\documentstyle[11pt,aaspp4]{article}

\slugcomment{To be published in Ap.J., Volume 522 No. 2 (1999 Sep 10).}
\lefthead{van der Tak et~al. }
\righthead{Temperature and density around massive YSO GL 2591}

\def\htwo{H$_2$}
\def\co{$^{13}$CO}
\def\cs{C$^{34}$S}
\def\water{H$_2$O}
\def\hcop{HCO$^+$}
\def\hhco{H$_2$CO}

\def\ammo{NH$_3$}

\def\gtsim{{_>\atop{^\sim}}}
\def\ltsim{{_<\atop{^\sim}}}
\def\vlsr{v$_{\rm LSR}$}
\def\kms{km~s$^{-1}$}

\def\rcm{cm$^{-1}$}
\def\scm{cm$^{-2}$}
\def\ccm{cm$^{-3}$}
\def\mic{$\mu$m}
\def\klm{k$\lambda$}
\def\chisq{$\chi^2$}
\def\msol{M$_{\odot}$}
\def\lsol{L$_{\odot}$}

\def \apj #1 #2 {ApJ, #1, #2}
\def \apjl #1 #2 {ApJ, #1, L#2}
\def \apjs #1 #2 {ApJS, #1, #2}
\def \aap #1 #2 {A\&A, #1, #2}
\def \aal #1 #2 {A\&A, #1, L#2}
\def \aas #1 #2 {A\&AS, #1, #2}
\def \aj #1 #2 {AJ, #1, #2}
\def \mnras #1 #2 {MNRAS, #1, #2}
\def \pasp #1 #2 {PASP, #1, #2}
\def \araa #1 #2 {ARA\&A, #1, #2}

\begin{document}
\title{The Impact of the Massive Young Star GL 2591 on its Circumstellar
Material: Temperature, Density and Velocity Structure }

\author{Floris F. S. van der Tak \& Ewine F. van Dishoeck}
\affil{Sterrewacht, P.O. Box 9513, 2300 RA Leiden, The Netherlands}
\authoraddr{Sterrewacht, P.O. Box 9513, 2300 RA Leiden, The Netherlands}
\author{Neal J. Evans II\altaffilmark{1} \& Eric J. Bakker\altaffilmark{1}} 
\affil{Department of Astronomy, University of Texas, Austin, TX 78712}
\and
\author{Geoffrey A. Blake}
\affil{Division of Geological and Planetary Sciences, California
Institute of Technology, MS 150--21, Pasadena, CA 91125} 

\altaffiltext{1}{Visiting Astronomer, Kitt Peak National Observatory, National Optical
Astronomy Observatories, which is operated by the Association of Universities
for Research in Astronomy, Inc. (AURA), under cooperative agreement with the
U.S. National Science Foundation.}

\begin{abstract}
\label{s:abs}
  
The temperature, density and kinematics of the gas and dust
surrounding the luminous ($2\times 10^4$~\lsol) young stellar object
GL~2591 are investigated on scales as small as
$\sim 100$~AU, probed by $4.7$~\mic \ absorption spectroscopy, to over
$60,000$~AU, probed by single--dish submillimeter spectroscopy.  These
two scales are connected by interferometric $86-115$~and $226$~GHz
images of size $30,000$~AU and resolution $2000$~AU in continuum and
molecular lines. The data are used to constrain the physical structure
of the envelope and investigate the influence of the young star on its
immediate surroundings.
  
  The infrared spectra at $\lambda / \Delta\lambda\approx 40,000$
  indicate an LSR~velocity of the \co\ rovibrational lines of $-5.7
  \pm 1.0$~\kms, consistent with the velocity of the rotational lines
  of CO.  In infrared absorption, the $^{12}$CO lines show wings out
  to much higher velocities, $\approx -200$~\kms, than are seen in
  the rotational emission lines, which have a total width of $\approx
  75$~\kms. This difference suggests that the outflow seen in
  rotational lines consists of envelope gas entrained by the ionized
  jet seen in Br$\gamma$ and [\ion{S}{2}] emission.  The outflowing
  gas is warm, $T >100$~K, since it is brighter in CO $J=6\to 5$ than
  in lower-$J$ CO transitions.
  
  The dust temperature due to heating by the young star has been
  calculated self-consistently as a function of radius for a power-law
  density distribution $n=n_0 r^{-\alpha}$, with $\alpha=1-2$.  The
  temperature is enhanced over the optically thin relation $(T\sim
  r^{-0.4})$ inside a radius of $2000$~AU, and reaches $120$~K at
  $r\ltsim 1500$~AU from the star, at which point ice mantles should
  have evaporated. The corresponding dust emission can match the
  observed $\lambda \geq 50$ \mic\ continuum spectrum for a wide range
  of dust optical properties and values of $\alpha$. However,
  consistency with the C$^{17}$O line emission requires a large dust
  opacity in the submillimeter, providing evidence for grain
  coagulation.  The $10-20$~\mic\ emission is better matched using
  bare grains than using ice-coated grains, consistent with
  evaporation of the ice mantles in the warm inner part of the
  envelope.  Throughout the envelope, the gas kinetic temperature as
  measured by H$_2$CO line ratios closely follows the dust
  temperature.
  
  The values of $\alpha$ and $n_0$ have been constrained by modelling
  emission lines of CS, HCN and \hcop\ over a large range of critical
  densities. The best fit is obtained for $\alpha=1.25 \pm 0.25$ and
  $n_0=(3.5 \pm 1)\times 10^4$ cm$^{-3}$ at $r=30,000$~AU, yielding an
  envelope mass of $(42\pm 10)$~\msol\ inside that radius.  The
  derived value of $\alpha$ suggests that part of the envelope is in
  free-fall collapse onto the star.  Abundances in the extended
  envelope are $5\times 10^{-9}$ for CS, $2\times 10^{-9}$ for
  H$_2$CO, $2\times 10^{-8}$ for HCN and $1\times 10^{-8}$ for
  HCO$^+$. The strong near-infrared continuum emission, the Br$\gamma$
  line flux and our analysis of the emission line profiles suggest
  small deviations from spherical symmetry, likely an evacuated
  outflow cavity directed nearly along the line of sight. The
  $A_V\approx 30$ towards the central star is a factor of 3 lower than
  in the best-fit spherical model.
  
  Compared to this envelope model, the OVRO continuum data show excess
  thermal emission, probably from dust.  The dust may reside in an
  optically thick, compact structure, with diameter $\ltsim 30$~AU and
  temperature $\gtsim 1000$~K, or the density gradient may steepen
  inside $1000$~AU.  In contrast, the HCN line emission seen by OVRO
  can be satisfactorily modelled as the innermost part of the power
  law envelope, with no increase in HCN abundance on scales where the
  ice mantles should have been evaporated. The region of hot, dense
  gas and enhanced HCN abundance ($\sim 10^{-6}$) observed with the
  {\it Infrared Space Observatory} therefore cannot be accommodated as an
  extension of the power-law envelope. Instead, it appears to be a
  compact region ($r<175$~AU, where $T>300$~K) where high-temperature
  reactions are affecting abundances. 

\end{abstract}


\section{Introduction} 
\label{s:intro}

The dust and molecular gas around low-mass young stellar objects
(YSOs)\footnote{We will use the term ``Young Stellar Object'', or YSO,
  for a gas and dust cloud that derives the bulk of its luminosity
  from nuclear burning, but that is still embedded in a molecular
  cloud, as opposed to ``protostar'', an object primarily radiating
  dissipated gravitational energy. Stellar objects are ``massive'' if
  they emit a substantial Lyman continuum.}  is observed to consist of
a disk of typical size $10^2$~AU, and a spherical power law envelope
extending out to $r \gtsim 10^4$~AU (0.05 pc). Perpendicular to the
disk, a bipolar molecular outflow is often seen, reaching velocities
up to $\sim 100$~\kms\ (see reviews by Shu \markcite{shu97}1997, Blake
\markcite{gab97}1997). Much less is known about the physical structure
around high-mass YSOs (Churchwell \markcite{chu93}1993,
\markcite{chu99}1999).  Because the formation of massive stars occurs
over much shorter time scales and involves much higher luminosities,
differences in the structure of the circumstellar environment may be
expected.  Submillimeter continuum observations have revealed the
presence of $\sim 100-1000$~\msol\ of dust and gas around massive
young stars (Walker, Adams \& Lada \markcite{wal90}1990; Henning,
Chini \& Pfau \markcite{hen92} 1992; Sandell\markcite{san94a} 1994;
Hunter, Phillips \& Menten \markcite{hun97}1997).  However, little
information exists about the distribution of this material within the
$(20-30)''$ single dish beams.  Angular momentum in a collapsing cloud
should produce a rotating disk, and the magnetic field is expected to
lead to the formation of a larger flattened structure.  Numerical
simulations indicate that such disks do indeed form (Yorke,
Bodenheimer \& Laughlin\markcite{yor95} 1995; Boss\markcite{bos96}
1996), but observational evidence for disks around massive young stars
has been sparse. Around Orion IRc2, an $\approx 20$~\msol\ star (Genzel
\& Stutzki\markcite{gen89} 1989) ---the best studied case by far---
less than $\sim 0.1$~\msol\ of neutral material resides in a disk
(Plambeck et~al.~\markcite{pla95}1995, Blake et
al.~\markcite{gab96}1996).  This apparent discrepancy may be due to
evaporation of the disks by the stellar ultraviolet continuum
(Hollenbach et~al.  \markcite{hol94}1994, Richling \& Yorke
\markcite{ric97}1997), which disperses a $1$~\msol\ disk around a
$10$~\msol\ star in $\sim 10^6$~yr. Well before the actual destruction
of the disk, its dust continuum emission may be hidden behind
free-free emission from the dense, ionized, evaporative flow, which
remains optically thick up to very high radio frequencies.

The large masses derived from the single-dish flux densities imply
that a physical model of the envelopes of massive young stars on $\sim
10^4$~AU scales is a prerequisite before any conclusions about the
smaller-scale structure can be drawn.  For these envelopes, four types
of models are found in the literature: homogeneous clouds of constant
density and temperature, inhomogeneous clouds with clumps but no
overall gradients (e.g., Wang et~al.~\markcite{wan93}1993; Blake et
al.~\markcite{gab96}1996), core-halo models (e.g., Little
et~al.~\markcite{lit94}1994) and power-law distributions (e.g., Carr
et~al.~\markcite{car95}1995).  The last category matches theoretical
considerations, which indicate density laws $\sim r^{-\alpha}$ with
$\alpha=2.0$ for clouds if they are thermally supported against
collapse and $\alpha=1.0$ if the support is nonthermal; clouds in
free-fall collapse should have $\alpha=1.5$ (Lizano \& Shu
\markcite{liz89}1989; Myers \& Fuller \markcite{mye92}1992; McLaughlin
\& Pudritz \markcite{mcl97}1997).  In addition, as a central star
develops, the material will not remain isothermal, and a combination
of thermal pressure, radiation pressure, and a stellar wind will stop
the infall process.  This may produce a shell of dense gas,
effectively flattening the average density law.

In this paper, we investigate the applicability of such models to
GL~2591, a site of massive star formation in the Cygnus~X region.
While most massive stars form in clusters, GL~2591 provides one of the
rare cases of a massive star forming in relative isolation, which
allows us to study the temperature, density and velocity structure of
the circumstellar envelope without confusion from nearby objects.
Although invisible at optical wavelengths, GL~2591 is very bright in
the infrared. Photometry over the full $2-200$~\mic\ range was
obtained by Lada et~al.~\markcite{lad84}(1984). Assuming a distance of
$1$~kpc, the luminosity is $\sim 2\times 10^4$~\lsol, leading to an
estimated stellar mass of 10~\msol. The infrared source is associated
with a weak radio continuum source (Campbell \markcite{cam84}(1984)
and with a powerful bipolar molecular outflow $>1'$ in extent (Lada
et~al.~\markcite{lad84}1984; Mitchell, Hasegawa \& Schella
\markcite{mit92}1992;Hasegawa \& Mitchell \markcite{has95}1995).

The distance to GL~2591 is highly uncertain. First, the source has no
optical counterpart, impeding a spectrophotometric determination.
Wendker \& Baars \markcite{wen74}(1974) associate GL~2591 with the
nearby ($\sim 1^\circ$), optically bright \ion{H}{2} region IC 1318~c,
for which Dickel et~al.~\markcite{dic69}(1969) determined a distance of
1.5~kpc. Second, the Galactic longitude of GL~2591 is close to
$90^\circ$, so that the Galactic differential rotation is almost
parallel to the line of sight, and the kinematic distance is poorly
constrained, $(4 \pm 2)$~kpc, assuming $R_0=8.5$~kpc and
$\Theta_0=220$~\kms.  The source may be a member of the Cyg OB2
association, in which case the distance is 2 kpc. Dame \& Thaddeus
\markcite{dam85}(1985) give the distance to Cygnus~X as 1.7~kpc, the
mass-weighted average of CO clouds in the region. The spread between
these clouds is large, $0.5-2.0$~kpc, which is probably real since the
line of sight is down a spiral arm.  Recent work on GL~2591 generally
assumes 1~kpc which provides a convenient scaling.  We adopt this
practice, but we will discuss how our conclusions are modified if the
distance is increased to 2~kpc.

The large columns of dust and gas toward GL~2591 block our view of the
stellar photosphere, but give rise to a high luminosity infrared
source, which allows detection of infrared absorption lines in the
colder foreground material. Indeed, one of our main motivations for
studying this source is the possibility for sensitive complementary
infrared data from the ground and from space, in particular with the
{\it Infrared Space Observatory} (ISO). Previous observations of CO
and \co\ $4.7$~\mic\ absorption lines by Mitchell
et~al.~\markcite{mit89}(1989) suggested a cold ($38$~K) and a hot
($1010$~K) component in the quiescent gas, as well as a blueshifted
warm ($\sim 200$~K) component. Carr et al.~\markcite{car95}(1995)
detected absorption lines of C$_2$H$_2$ and HCN at $14$~\mic, implying
a density of $\approx 3\times 10^7$~\ccm\ for the warm and hot
components, and abundances of HCN that are a factor of $\sim 100$ higher
than those in the extended envelope. Recent ISO observations with the
{\it Short Wavelength Spectrometer} (SWS) have resulted in the
detection of hot ($\sim 300$~K), abundant gas-phase \water\ (Helmich
et~al.~\markcite{fph96}1996; van Dishoeck \& Helmich
\markcite{evd96a}1996).  Even higher temperatures ($\sim 1000$~K) are
seen in the ISO $14$~\mic\ absorption profiles of C$_2$H$_2$ and HCN
(Lahuis \& van Dishoeck \markcite{lah97}1997).

Carr et~al.~\markcite{car95}(1995) used millimeter observations of CS,
HCN, HCO$^+$ and their isotopes to constrain the density structure of
a one-dimensional power law envelope model. They ruled out
constant-density models and found $\alpha=1.5$ to fit the data
somewhat better than $\alpha=1.0$ or $2.0$.  Comparison of the
strength of HCN millimeter emission and infrared absorption lines led
them to conclude that the size of the dense, hot region is $\ltsim
3''$. However, it was not clear how this inner region relates to the
large scale structure, especially since the velocities of the infrared
lines of CO appeared to differ by $5-10$~\kms\ from those of the
rotational lines (Mitchell et al. 1989).  

In this paper, we present new single-dish submillimeter data,
millimeter interferometry and infrared absorption line observations of
GL 2591, thereby extending the previous data in several ways. First,
the interferometer probes $\sim 10$ times smaller scales than the
single-dish observations, bridging the gap toward the infrared
absorption, which occurs in a pencil beam set by the size of the
emitting region. The imaging capability enables us to relate the
interferometer data to the single-dish submillimeter data.  Second,
the new single-dish data cover a larger frequency range
($86-650$~GHz), while also probing higher gas densities (up to
$10^8$~\ccm) and temperatures (up to $200$~K) than the observations
presented by Mitchell et~al.~\markcite{mit92}(1992) and by Carr
et~al.~\markcite{car95}(1995). The new infrared spectra are at
slightly higher resolution, $\lambda / \Delta \lambda \approx 40,000$,
and have lower noise than previous data by Mitchell
et~al.~\markcite{mit89}(1989) and can thus help to resolve the
discrepancy between the infrared and millimeter velocities. The
combined data will be used to constrain the physical and kinematical
structure of the dust and molecular gas on scales of $\sim100$ to
$\sim30,000$~AU. The chemical composition of the envelope will be
discussed in a subsequent paper.

This paper is organized as follows. First we present the observations
(\S~\ref{s:obs}) and their direct implications (\S~\ref{s:res}).  In
\S~\ref{s:phys}, we develop a model for the physical structure of the
circumstellar envelope. The temperature structure of the dust is
calculated in \S~\ref{s:dusto}, the masses from gas and dust tracers
are compared in \S~\ref{s:masses}, leading to an estimate of the
submillimeter dust opacity, and the density distribution in the
envelope is obtained in \S~\ref{s:alpha}.  Possible alternative models
and deviations from spherical symmetry are discussed in
\S\S~\ref{s:ch} and~\ref{s:2dm}.  This model is subsequently compared
to the interferometer continuum and HCN line observations in
\S~\ref{s:o.models}, to search for evidence of a more compact
component and changes in the HCN abundance on small scales.  The paper
concludes with a summary of the main findings (\S~\ref{s:concl}).

\section{Observations} 
\label{s:obs}
\subsection{Interferometer Observations} 
\label{s:ovro}

Interferometer maps of various lines and continuum at $86-226$~GHz
were obtained with the millimeter array of the {\it Owens Valley Radio
  Observatory} (OVRO)\footnote{The Owens Valley Millimeter Array is
  operated by the California Institute of Technology under funding
  from the U.S. National Science Foundation (AST96-13717).}. The OVRO
interferometer consists of six 10.4~m antennas on North-South and
East-West baselines. Three frequency settings were observed, the basic
parameters of which are listed in Table~\ref{ovro_log.tab}. This
paper only presents the continuum, CO and HCN (+isotopic) data; the SO,
SO$_2$ and CH$_3$OH results will be discussed in a future paper.

The gains and phases of the antennas were monitored with snapshots of
the quasars 2023+336 and 2037+511; the bandpass was checked against 3C273,
3C454.3 and 3C111. Due to the high level of atmospheric decorrelation
at $226$~GHz, it was not possible to impose phase closure, as was the
case for the $86-115$~GHz data. Instead, approximate phase solutions
were derived by setting the phase of the calibrator to zero on every
baseline. Absolute flux calibration is based on 15-minute integrations
on Uranus and Neptune. The flux densities at a given frequency found
for the phase calibrator on different days agree to within the
estimated calibration uncertainty of $10$~\%. At 226~GHz, the likely
uncertainty is closer to $20$ \% due to the higher atmospheric phase
noise. Data calibration was performed using the MMA package (Scoville
et~al.~\markcite{sco93}1993); further analysis of the OVRO data was
carried out within MIRIAD.

\subsection{Single Dish Submillimeter Observations}
\label{s:submm}

Most of our single-dish submillimeter data were obtained with the 15-m
James Clerk Maxwell Telescope (JCMT)\footnote{The James Clerk Maxwell
  Telescope is operated by the Joint Astronomy Centre, on behalf of
  the Particle Physics and Astronomy Research Council of the United
  Kingdom, the Netherlands Organization for Scientific Research and
  the National Research Council of Canada.} on Mauna Kea, Hawaii
during various runs in 1995 and 1996. The antenna has an approximately
Gaussian main beam of FWHM $18''$ at 230 GHz, $14''$ at 345~GHz, and
$11''$ at 490 GHz. Detailed technical information about the JCMT and
its receivers and spectrometer can be found in Matthews
\markcite{mat95}(1995), or on-line at {\tt
  $<$http://www.jach.hawaii.edu/JCMT/home.html$>$}. Receivers A2, B3i
and C2 were used as front ends at 230, 345 and 490 GHz, respectively.
The Digital Autocorrelation Spectrometer served as the back end, with
continuous calibration and natural weighting employed. To subtract the
atmospheric and instrumental background, a reference position
$180''$~East was observed, except for the CO lines, where an $1800''$
offset was used. Values for the main beam efficiency $\eta_{mb}$,
determined by the JCMT staff from observations of Mars and Jupiter,
are $0.69$, $0.58$ and $0.53$ at 230, 345 and 490~GHz for the 1995
data, and $0.64$, $0.60$ and $0.53$ for 1996. Absolute calibration
should be correct to $20\%$, except for data in the 230~GHz band from
May 1996, which have an uncertainty of $\approx 50\%$ due to technical
problems with receiver A2. Pointing was checked every 2 hours during
the observing and was usually found to be within $2''$ and always
within $4''$. Integration times are 30-40 minutes per frequency
setting, resulting in rms noise levels in T$_{\rm mb}$ per $625$~kHz
channel ranging from $\approx 30$~mK at 230~GHz to $\approx 100$~mK at
490~GHz.  A small $104'' \times 104''$ map was made using the {\it
  on-the-fly} mapping mode in the $^{13}$CO 3-2 line, with an rms
noise of 2~K~\kms.

Single dish observations of molecular lines in the $86-115$~GHz range
were made in October and November of 1995 with the NRAO~12m
telescope\footnote{The National Radio Astronomical Observatory is
  operated by Associated Universities, Inc., under contract with the
  U.S. National Science Foundation.} on Kitt Peak. The
receiver was the 3mm SIS dual channel mixer. For the back ends, one
256-channel filter bank was split into two sections of 128 channels at
100~kHz ($0.34$~\kms) resolution, and the Hybrid Spectrometer was
placed in dual channel mode with a resolution of 47.9~kHz
($0.16$~\kms).  The beam width is $63''$~FWHM and the main beam
efficiency $\eta_{mb}=0.86$.  Pointing is accurate to $10''$ in
azimuth and $5''$ in elevation.

Observations of the CO and \co\ $J=6\to 5$ lines near 650~GHz were
carried out in May 1995 with the 10.4-m antenna of the Caltech
Submillimeter Observatory (CSO)\footnote{The Caltech Submillimeter
  Observatory is operated by the California Institute of Technology
  under funding from the U.S. National Science Foundation
  (AST96-15025)}. The back ends were the Acousto-Optical Spectrometers
(AOS) with $500$~MHz and 50 MHz bandwidth.  At $650$~GHz, the CSO has
a beam size of $11''.2$~FWHM and a main beam efficiency
$\eta_{mb}=0.40$. Pointing is accurate to $4''$. All single-dish data
were reduced with the IRAM CLASS package.

\subsection{Infrared Observations}
\label{s:phx}

Spectra of GL~2591 near the CO $v=1\gets 0$ band at $4.7$~\mic\ were
obtained with the Phoenix spectrometer mounted at the f/15 focus of
the NOAO 2.1m telescope on Kitt Peak. A single grating order is
projected onto an InSb array with 1024 pixels in the dispersion
direction, covering a $1500$~\kms\ bandpass at a resolution of
$\lambda/\Delta \lambda \approx 40,000$ ($7.5$~\kms). Technical
information about the instrument can be found at {\tt
  $<$http://www.noao.edu/kpno/phoenix/phoenix.html$>$}.  Three
wavelength regions have been observed: near $2155$~\rcm\ on 1997
April~1, near $2112$~\rcm\ on 1997 October~23 and near $2134$~\rcm\ on
1997 October~24. On-source integration times were 1~hour for each
wavelength setting, spread over 90-second scans. The weather was
partly cloudy during all nights, with a high and variable humidity.

Reduction was carried out with the NOAO IRAF package. Consecutive
array frames were subtracted from each other to remove instrumental
bias and (to first order) the atmospheric background, but the high
variability of the background required a second correction
during the aperture extraction. A dome flat field was used to correct
for sensitivity variations across the chip.  Wavelength calibration is
based on the telluric CO lines in the spectrum of a reference object,
which was the Moon in April and Vega in October.  To remove the
telluric CO and \co\ lines, the source data are divided by the
standard star data, scaled to the appropriate air mass. The absorption
in telluric water lines varied significantly over a 1-hour exposure,
and scale factors for the cancellations of these features were
determined empirically in order to obtain a straight continuum. For
the April observations, the cancellation of telluric CO lines is
limited by the use of the Moon as reference object. Since this is an
extended source, it will illuminate the optics differently, leading to
a broadening of the telluric features. This broadening can give
spurious ``emission'' features when the source data are divided by the
reference data.

\section{Results}
\label{s:res}
\subsection{Interferometer Maps of 86-226 GHz Continuum}
\label{s:c.maps}

Figure~\ref{c_obs.fig} presents the continuum emission of GL~2591 at
87, 106, 115~and 226~GHz. These maps were produced from the OVRO data
in the standard way, using uniform weight in the Fourier transform and
deconvolution with the CLEAN algorithm. Self-calibration on the
brightest CLEAN components improved the phases of the $uv$ data. The
beam FWHM and noise level of the maps can be found in Table~1.

Two sources are detected, separated by $\sim 6''$. At 87~GHz, the
SW~source is the brightest, but at higher frequencies, the NE~source
begins to dominate the flux in the field. At 226~GHz, where the
sensitivity is lower, only the NE source is detected.  No sources were
detected in the high resolution array configuration at $226$~GHz,
which is why the beam size is similar to that of the lower-frequency
images.

The positions and flux densities of the sources were measured by
fitting models to the $uv$ data before and after self-calibration,
respectively. The simplest model that describes the data well consists
of a point source for the NE object and a Gaussian for the SW source.
The results are summarized in Table~\ref{posflux.tab}, together with
cm-wave measurements from the literature. Based on the positional
agreement to within $0\farcs2$, we associate the SW~source with radio
source~1 from Campbell \markcite{cam84}(1984), and the NE~source with
the infrared source from Tamura et~al.~\markcite{tam91}(1991) and
with radio source~3 from Campbell.

The SW~source has an approximately flat spectrum between 6.1~cm and
0.3~cm, with a spectral index $\gamma$ ($S_\nu \propto \nu^\gamma$)
between 5~GHz and 87~GHz of $-0.03 \pm 0.01$. This suggests free-free
emission from an optically thin \ion{H}{2} region, for which
$\gamma=-0.1$. The relation of the SW source to the molecular cloud
core will be discussed further in \S~\ref{s:2dm}.  For the NE source,
VLA observations (Campbell \markcite{cam84}1984; Tofani
et~al.~\markcite{tof95}1995) indicate a spectral index between
$6.1$~cm and $3.6$~cm of $\gamma \approx 0.6$, the value expected for
a spherical ionized wind.  Extrapolating along this spectral index
gives an $86$~GHz flux density of $3.6$~mJy, about an order of
magnitude below that observed with OVRO.  The spectral index of the
NE~source at millimeter wavelengths is somewhat uncertain because of
calibration problems at $226$~GHz.  Most of the dynamic range in the
presented $226$~GHz image was achieved by self-calibration; this
process may have falsely attributed the flux of the SW source to the
NE~source. We therefore regard the measured flux density of
$151.4$~mJy as an upper limit. A lower limit is $\approx 70$~mJy,
which holds if the SW~source has a constant spectral index up to
$226$~GHz, but such a low value is unlikely since the emission was
detected at the position of the infrared source.  The best value for
the spectral index of the NE~source is $1.7 \pm 0.3$. This value could
arise in an \ion{H}{2} region with a slight density gradient, which
raises the question how this gas is related to the ionized wind seen
at cm wavelengths.  The same combination of quiescent and expanding
components is seen in the surrounding neutral material (Mitchell
et~al.~\markcite{mit89}1989). This interpretation can be tested with
interferometric observations of radio recombination lines.  However,
the derived spectral index is also close to $2.0$, suggesting black
body emission.  Regardless of whether this emission is due to dust or
to ionized gas, the low brightness temperatures of only $1-2$~K at all
frequencies observed with OVRO imply that the emission fills only a
small fraction of the beam. We show in \S~\ref{s:c.models} that it is
not due to extended emission.

If all of the emission arises in an (ultra-)compact \ion{H}{2} region,
the absence of a spectral turnover up to $226$~GHz implies an emission
measure $\gtsim 10^{10}$~pc~cm$^{-6}$, a source diameter $\ltsim
20$~AU and an electron density $\gtsim 10^7$~\ccm.  With the
recombination rate in an optically thick \ion{H}{2} region (``case
B''), $\alpha_B = 2.59 \times 10^{-13}$~cm$^3$~s$^{-1}$ (Osterbrock
\markcite{ost91}1991), the stellar supply of Lyman continuum photons
is estimated to be $\gtsim 3\times 10^{45}$~s$^{-1}$. Using the lower
limit, the stellar atmosphere models by Thompson
\markcite{tho84}(1984) indicate an effective temperature of
$25,000$~K and a luminosity of $8000$~\lsol. More detailed models by
Schaerer \& de Koter \markcite{sdk97}(1997) suggest somewhat lower
values for $T_{\rm eff}$ and L/\lsol, but this difference may easily
be compensated for by the ``leaking out'' of ionizing photons.
Comparing this estimate of the stellar luminosity with the observed
infrared flux of GL~2591 by Lada et~al.~\markcite{lad84}(1984) leads
to a lower limit to the distance of $>0.6$~kpc.  Even this limit value
is highly uncertain since additional luminosity could be provided by
low-mass companion stars.

Alternatively, dust could be responsible for the emission, in which
case the characteristic temperature is significantly lower.  For
instance, for $T=100$~K, the source diameter would be $200$~AU.  Dust
emission on these spatial scales could arise in the innermost regions
of a power law envelope. However, we will show in \S~\ref{s:o.models}
that the OVRO continuum emission from GL~2591 is caused by a separate,
compact dust component after developing a model for the more extended
envelope from the single-dish observations in \S~\ref{s:phys}. It
should be noted that the continuum fluxes seen in the interferometer
are only a small fraction of those observed with single-dish antennas,
$\sim 5$\% at $226$~GHz: most of the envelope emission is filtered out
by the interferometer.

\subsection{Interferometer Maps of Molecular Line Emission}
\label{s:l.maps}

In Figure~\ref{l_obs.fig}, the OVRO images of the $J=1\to 0$ line of
CO, HCN, H$^{13}$CN and HC$^{15}$N are presented, after deconvolution
with CLEAN. For HCN and H$^{13}$CN, the integrated emission includes
the hyperfine components; C$^{17}$O was not detected. The CO map does
not include the flux on projected baselines shorter than $7$~\klm.
Also shown are spectra at the image maxima averaged over the central
beam area, obtained from deconvolved image cubes at full spectral
resolution and coverage, except that the outermost 8 channels in the wide
mode and 4 in the narrow mode were left out because of poor bandpass. The
self-calibration solutions from the continuum data, which have a
higher signal-to-noise ratio, were adopted for the line data. No
continuum was subtracted. Beam size and rms noise of the images can be
found in Table~1.

Due to lower signal-to-noise, the positional accuracy of the line data
is lower than that of the continuum, but for all lines, the emission
maximum lies within half a beam from the NE continuum source.  Hence,
the molecular line emission is associated with the infrared source,
not with the SW continuum source. The emission appears compact, except for
HCN, which is elongated, and CO, which, although very bright, is not
centrally concentrated like the other molecules. Instead, the CO line
appears to trace irregularities in the molecular cloud surrounding the
star-forming core and/or in the outflow.  

The right-hand panels of Figure~\ref{l_obs.fig} show the profiles of
the HCN, H$^{13}$CN and HC$^{15}$N $J=1\to 0$ lines observed with the
NRAO~12m telescope. Also plotted, as thick lines, are the OVRO
spectra, constructed from the maps in the left-hand panels after
convolution to the resolution of the 12m.  Only $\approx 20$~\% of the
total line flux is recovered with OVRO, although the two telescopes
are seen to trace kinematically the same gas.  The
missing flux in the interferometer data ($\approx 80$~\%) must be on
angular scales of $\gtsim 30''$, to which the array is not sensitive.
The CO emission, after convolution to a $63''$~beam, has a peak
brightness of $\approx 1.1$~K, or about 5~\% of the value measured by
Lada et al.~\markcite{lad84}(1984) at the NRAO~12m. Thus, $95$~\% of
the CO $J=1\to0$ emission arises on $\gtsim 30''$ scales. 

\subsection{Single-dish Molecular Line Emission} 
\label{s:emlines}

Table~\ref{obs_submm.tab} summarizes the results of our single-dish
observations. All spectral features show a Gaussian component at
\vlsr~$=(-5.5\pm 0.2)$~\kms\ with an FWHM of $(3.3 \pm 0.6)$~\kms,
while several lines show additional blueshifted emission, which has
been fitted by a second Gaussian component at $(-6.9\pm 0.7)$~\kms\ 
with an FWHM of $(6.2 \pm 2.5)$~\kms. The profiles of the CO $J=3\to
2$ and $J=4\to 3$ lines can be reproduced by the same combination of
two emission components, but two additional absorption components at
\vlsr$=0.0$ and $-8.0$~\kms\ are required. These absorptions are also
seen in the OVRO CO and HCN $J= 1\to 0$ line profiles.  The absorption
at zero \vlsr\ is due to an extended cold foreground cloud, seen in
emission at $\gtsim 60''$ offsets (Mitchell
et~al.~\markcite{mit92}1992), while the absorption at $-8$~\kms\ may
be intrinsic to the source. Since neither of these absorption features
is present in the CO $J=6\to5$ line, the absorbing gas must be cold
and/or tenuous.

The outflow is observed in the broad wings of the $^{12}$CO emission
line profiles, which can be described with a Gaussian of FWHM
$=20$~\kms, centered at \vlsr~$=-7$~\kms. The flow is also seen as the
broad emission components in lines of CS, \hhco, HCN, \hcop\ and
H$^{13}$CO$^+$.  The high-velocity CO is brighter in the $J=6\to 5$
than in the $J=4\to 3$ line, both measured in an $11''$ beam, implying
a kinetic temperature of at least 100~K on a scale of $10^4$ AU.  The
large extent of the flow (Lada et al.~\markcite{lad84}1984; Mitchell
et al.~\markcite{mit92}1992) may explain why the gas seen in infrared
absorption at $-21.5$ \kms\ (\S~\ref{s:irvel}) is not prominent in the
OVRO data: the interferometer filters out most of the emission. The
fast molecular jet at \vlsr$=-50$ to $-200$~\kms\ was not covered by
the OVRO spectrometers.

Figure~\ref{jcmt_maps.fig} shows the \co\ $J=3\to 2$ map of GL~2591,
both in the line wings and integrated over the entire profile. Also
shown are $450$~and $850$~\mic\ continuum observations with the
Submillimeter Common User Bolometer Array (SCUBA) at the JCMT,
provided by M. van den Ancker and G. Sandell (private communication).
The lowest contour drawn for the \co\ wing emission, $5$~K~\kms, is
the brightness expected for the molecular cloud surrounding the
star-forming region, assuming T$_{\rm kin}=15$~K, $n=10^4$~\ccm,
N(\co)$=1.7\times 10^{16}$~\scm\ ($N({\rm H}_2)=10^{22}$~\scm) and
$\Delta V =2$~\kms.  The gas and dust tracers agree reasonably well on
the extent of the dense molecular cloud core. The diameter at which
the intensity has dropped to half its peak value is $\approx 20''$ in
both continuum maps, measured both N-S and E-W. The \co\ diameter is
closer to $\approx 30''$, but the map of the line wings demonstrates
that this elongation is due to the outflow.  Data from the KAO
scanning photometer at $50$~and $100$~\mic\ (P.\ Harvey, private
communication) indicate a source diameter of $\approx 20''$.  Based on
these observations, we conclude that the star-forming core is confined
to a region of radius $\approx 30''$, or $30,000$~AU at 1~kpc. This
estimate is accurate to a factor of 2 or so, sufficient to be a useful
constraint for the models to be developed in \S~\ref{s:phys}. Although
the circumstellar envelope may extend to larger radii, the density and
temperature drop to levels that produce little emission in most of the
tracers used in this study.

\subsection{Infrared Lines and Velocity Structure}
\label{s:irvel}

Figure~\ref{phxdata.fig} presents the calibrated infrared spectra. The
apparent emission features in the $2155$~\rcm\ setting are due to the
use of the Moon as the reference object (see \S~\ref{s:phx}). All
absorption features were identified as rovibrational lines of CO and
\co\ using frequencies calculated from the Dunham coefficients of
Farrenq et al.~\markcite{far91}(1991). In the $2155$~\rcm\ setting,
which has the highest signal-to-noise ratio, lines of vibrationally
excited CO are detected, which are marked with CO* in
Figure~\ref{phxdata.fig}.  Table~\ref{irvel.tab} lists the center
velocities, FWHM line widths and equivalent widths of all the detected
lines, measured by fitting Gaussian profiles to the data. The
uncertainty in the line parameters are dominated by errors in
continuum placement and by the effects of imperfect cancellation of
telluric features. 

The $^{12}$CO fundamental band profiles are very broad and asymmetric,
which prohibits the fitting of a Gaussian or Voigt profile. The
strongest absorption occurs at \vlsr~$=(-21.5 \pm 2.6)$~\kms, at which
position \co\ absorption is also detected.  From spectra covering many
more lines than this work but at a lower velocity resolution, Mitchell
et al.~\markcite{mit89}(1989) derived a temperature of $200$~K and a
CO column density of $6.6\times 10^{18}$~\scm\ for this gas component,
which they called a ``$-28$~\kms\ component''. The line width measured
here, $\approx 22$~\kms, is larger than the value of $12.5$~\kms\ by
Mitchell et al., so that the column density may also be higher.  Our
limited line coverage does not allow an accurate measurement of the
column density.  This component may be related to the outflow seen in
the CO rotational lines, for which a temperature of $\gtsim 100$~K was
found in \S~\ref{s:emlines}.

The shape of the broad blueshifted wings of the $^{12}$CO $v=1\gets 0$
lines suggests an origin in a wind. The position where the continuum
is reached is measured to be $(-196 \pm 3)$~\kms, which gives a
terminal velocity of $(173 \pm 3)$~\kms.  This value is much lower
than the typical terminal velocity of the winds of early B~stars
measured from ionic ultraviolet lines, $\sim 1500$~\kms\ (Lamers, Snow
\& Lindholme\markcite{lam95} 1995), but much higher than the
velocities seen in CO rotational emission in this source (Choi, Evans
\& Jaffe \markcite{cho93}1993; \S~\ref{s:emlines}) or in fact in any
source in CO rotational emission (Choi et al.~\markcite{cho93}1993;
Shepherd \& Churchwell \markcite{dss95}1995, \markcite{dss96}1996).  A
velocity of $\approx 200$~\kms\ is comparable to the velocity of
$500$~\kms\ seen in the [\ion{S}{2}] $\lambda6731$ line by Poetzel,
Mundt \& Ray \markcite{poe92}(1992), and to the total width of
\ion{H}{1} infrared recombination lines in objects similar to GL~2591
(Bunn et al. \markcite{bun95}1995). The range of velocities seen in
$^{12}$CO absorption suggests an origin in envelope gas entrained by
the ionized jet.  Figure~\ref{ir-submm.fig} compares the CO and \co\ 
infrared line profiles with those observed in rotational emission with
the JCMT and the CSO. 

The \co\ absorption lines show a second feature at a less negative
\vlsr. This gas was also seen by Mitchell et
al.~\markcite{mit89}(1989), who attributed it to the quiescent
molecular cloud core out of which the star formed. However, Mitchell
et al.  measured a \vlsr\ of $-8$~to $-11$~\kms\ for this component,
which does not agree with the velocity of rotational lines.  We
measure \vlsr=$(-5.7\pm 1.0)$~\kms, consistent with the velocity of
the millimeter emission lines of CO, CS and other molecules, $(-5.5
\pm 0.2)$~\kms\ (\S~\ref{s:emlines}).  This value was determined from
the R(3), R(10) and R(16) lines, which show no contamination by
telluric features.  Observations of the $v=2\gets 0$ band of CO with
Phoenix give velocities similar to those found here (C.~Kulesa,
private communication).  The line width is measured to be $\approx
10$~\kms, similar to that found by Mitchell et al., but a factor of
$2-3$ larger than the width of the rotational lines in this source
(\S~\ref{s:emlines}).  Because this linewidth is close to our
resolution, it may be an overestimate.  Thus, we associate the
submillimeter emission and infrared absorption lines with the same
gas.

\section{Physical Structure of the Envelope}
\label{s:phys}

In order to analyze the interferometer line and continuum data, a good
physical model of the extended envelope is a prerequisite. The first
step will be to constrain the temperature structure of the
circumstellar envelope, while the second step is to determine the
density structure.

\subsection{Dust Temperature Structure}
\label{s:dusto}

The dust temperature in the envelope as a function of distance from
the star was calculated with the computer program by Egan, Leung \&
Spagna \markcite{ega88}(1988).  The number density of dust grains was
assumed to follow an $r^{-\alpha}$ power law. Initially, trial values
of $0.5-2.0$ were used for $\alpha$, but anticipating the constraints
on $\alpha$ provided by the CS line emission (\S~\ref{s:alpha}), we
only present results for $1.0<\alpha<1.5$.  The inner radius was fixed
at $200$~AU ($0\farcs2$ at 1~kpc), much smaller than the OVRO beam and
small enough that it does not influence the calculated brightness.
Based on the maps presented in \S~\ref{s:emlines}, we used $30,000$~AU
as the outer radius of the models. Larger values for the outer radius
give the same conclusions. The observed luminosity of $2\times
10^4$~\lsol\ is used, and it is assumed that its source is a zero age
main sequence star, in which case the effective temperature is
$\approx 30,000$~K (Thompson \markcite{tho84}1984; Schaerer \& de
Koter \markcite{sdk97}1997; Hanson, Howarth \& Conti
\markcite{han97}1997). However, varying $T_{\rm eff}$ by a factor 2 at
constant $L$ (hence varying the stellar radius) changes the computed
submillimeter flux density by less than 1\%. The stellar spectral type
and any contribution from accretion to the luminosity thus cannot be
derived from these models.

The program solves self-consistently for the grain heating and cooling
as a function of radius. The model consists of 100 concentric
spherical shells, roughly logarithmically spaced. No external
radiation field was applied. Comparison to observations proceeds by
integrating the radiative transfer equation on a grid of lines of
sight followed by convolution to the appropriate beam (Butner
et~al.~\markcite{but91}1991).  The only free parameter in this process
is the number density of dust grains at a given radius, or
equivalently the total dust mass. This parameter was constrained by
JCMT $450$, $850$, and $1100$~\mic\ fluxes provided by G.~Sandell
(private communication), obtained in the same manner as described in
Sandell \markcite{san94a}(1994). At these long wavelengths, the dust
emission is optically thin and least prone to geometrical effects. In
addition, the beams are small enough ($\approx 18''$~FWHM) that only
the dense core is probed.  The model is also compared to $60$, $95$,
$110$~and $160$~\mic\ photometry from Lada
et~al.~\markcite{lad84}(1984), observed with the KAO in a 49$''$ beam,
and $2-20$~\mic\ data from Aitken et~al.~\markcite{ait88}(1988),
obtained at UKIRT in a 4\farcs2 beam.

Initially, four sets of optical properties were considered for the
dust: the models for diffuse cloud dust by Draine \& Lee
\markcite{dra84}(1984) and Li \& Greenberg \markcite{lig97}(1997), and
for dense cloud dust by Mathis, Mezger \& Panagia
\markcite{mat83}(1983) and Ossenkopf \& Henning
\markcite{oss94}(1994). We use Ossenkopf \& Henning's model grains
with thin ice mantles coagulating at a density of $10^6$~\ccm, called
OH5 (column~5 of their Table~1). At visual and ultraviolet
wavelengths, the opacities are taken from Model~1 from Pollack
et~al.~\markcite{pol94}(1994), which is similar in spirit. The values
of the opacity and the albedo at these short wavelengths do not affect
the calculated spectrum since all light is absorbed close to the star.
We used a radius of $0.1$~\mic\ for the grains, and a mass density of
$2.3$~g~\ccm, the average of the values for silicate and amorphous
carbon. We neglect the effect of a distribution of grain sizes, which
would cause a range of dust temperatures at a given radius of $\sim
30$~\% or less (Wolfire \& Cassinelli \markcite{wol86}1986). Our
temperature profiles therefore represent averages over grain
populations, as in Churchwell, Wolfire \& Wood \markcite{cww90}(1990).

In Figure~\ref{sed.fig}, the calculated dust temperature profiles are
presented, with the curve $T=T_0 (r/r_0)^{-0.4}$ expected for optically thin
dust superposed. Here, $T_0$~is the temperature at the outer radius,
$28.7$~K. It is seen that inside a radius of $\approx 2000$~AU, the
dust temperature lies above the optically thin curve, and becomes
weakly dependent on the grain model and on $\alpha$. The location of
the outer edge of the model does not influence the results: for
instance, doubling the outer radius changes the calculated temperature
at a given radius by less than $0.5$~K, and the emergent flux
densities by less than $5$~\%.  The resulting infrared continuum
spectra are shown in the top part of Figure~\ref{sed.fig}.  Observed
and calculated submillimeter flux densities are summarized in
Table~\ref{phot_obs.tab}, along with the implied dust masses inside 30,000
AU and the $20$~\mic\ flux densities. The masses, or equivalently the
values of $n_0$, were chosen such that the best fit to the JCMT data
is obtained.  The same models reproduce the KAO data at 60, 95 and 110
$\mu$m to within a factor of $\sim 2$. In contrast, only the models
with Mathis et al.  dust match the near-infrared data, and none of the
models match the interferometric observations at 3000~\mic. These
discrepancies are discussed in \S\S~\ref{s:2dm} and~\ref{s:c.models}.

Close to the star, the calculated dust temperature is high enough
($\gtsim 100$~K) for the ice mantles to evaporate off the grains. In
the model with $\alpha=1.5$, half of the material along the central
line of sight is at temperatures $> 150$~K. Evaporation of icy mantles
is consistent with the ISO observations of abundant gaseous \water\ 
and a ratio of gaseous to solid \water\ of unity (Helmich
et~al.~\markcite{fph96}1996, van Dishoeck
et~al.~\markcite{evd96b}1996).  Therefore we considered a fifth dust
model, the Ossenkopf \& Henning \markcite{oss94}(1994) opacities for
the case of bare grains after coagulation at $10^6$~\ccm\ (column~2 of
their Table~1), which is denoted by OH2 in Table~\ref{phot_obs.tab}.
Models with bare grains close to the star and icy grains at large
radii (as in Churchwell et al. \markcite{cww90}1990) require
iteration, because the dust temperature and the optical properties
become interdependent.  Our computer program is not set up to do this.
Instead, we present calculations with either type of grain throughout
the cloud, which provide limiting cases. It was found that, while the
far-infrared brightness is insensitive to the presence of grain
mantles, the calculated mid-infrared flux densities are higher using
bare grains than using ice-coated coagulated grains, and closer to the
observed values (see Table 5).  We interpret this result as strong
evidence for grain modification by stellar radiation.  This
interpretation is supported by the $11.2$~\mic\ polarization feature,
which is due to crystalline olivine grains produced by dust annealing
(Aitken et al.~\markcite{ait88}1988; Wright et
al.~\markcite{wri99}1999).

Although varying $n_0$ can make any of the grain models agree with the
observed submillimeter continuum emission for either value of
$\alpha$, the implied dust masses vary by factors of $2-100$. This
discrepancy reflects differences in the absolute value of the grain
opacity at submillimeter wavelengths between the various dust models. In
\S~\ref{s:masses}, it will be shown that only the models with the
Ossenkopf \& Henning optical properties are consistent with the masses
derived from the C$^{17}$O emission.

\subsection{Dust and Gas Masses: Measuring the Submillimeter Dust Opacity}
\label{s:masses}

Given the temperature structure of the dust, the density distribution
in the envelope can be constrained from the molecular line data.  The
line emission of GL~2591 was modeled with a new computer program using
the Monte Carlo approach, developed in Leiden by Hogerheijde \& van
der Tak \markcite{hog98}(in prep.). The new program was extensively
tested against the code by Choi et~al. \markcite{cho95}(1995), which
was used previously by Carr et~al. \markcite{car95}(1995) to model
line emission from GL~2591. The program solves for the local radiation
field after each photon propagation and then derives the populations
in each molecular energy level. After the level populations have been
obtained, line profiles are calculated by line-of-sight integration
followed by convolution to the appropriate beam. The model consisted
of 20 spherical shells, spaced logarithmically. As in the dust models,
the outer radius was taken to be $30,000$~AU. The inner radius was
chosen small enough as to not influence the results. In particular,
the maximum density must exceed the critical density of all modelled
lines.

Radial density profiles of the form $n_0(r/r_0)^{-\alpha}$ were
considered, with $\alpha$ between $1.0$ and $2.0$. It is assumed that
the gas is heated to the dust temperature by gas-grain collisions
throughout the envelope; detailed calculations by Doty \& Neufeld
\markcite{dot97}(1997; their model with $M_{0.1}=100$~\msol,
$L/L_\odot=10^4$) show that this assumption is valid within
$20,000$~AU from the star.  The maximum difference outside this radius
is $20$~\%, which we will ignore.  The temperature structure
calculated for the dust for the same value of $\alpha$ was used.  The
intrinsic (turbulent) line profile was taken to be a Gaussian with a
Doppler parameter ($1/e$ width) of $1.6$~\kms, independent of radius.

The following isotopic abundance ratios have been used (Wilson \& Rood
\markcite{wil94}1994): $^{12}$C/$^{13}$C=60, $^{16}$O/$^{17}$O=2500,
$^{16}$O/$^{18}$O=500, $^{32}$S/$^{34}$S=22, $^{14}$N/$^{15}$N=270.
For \hhco, the ortho/para ratio was fixed at 3. The rate coefficients
for collisional de-excitation as listed by Jansen et al.
\markcite{jan94}(1994) and Jansen \markcite{jan95}(1995) have been
used.  No rate coefficients are available for $T\gtsim 200$~K, except
for HCN and CO, and it is assumed that the collisional de-excitation
rate coefficients become temperature-independent at high temperatures.

The values of $n_0$ found for a given $\alpha$ from the dust emission
using various opacity curves were tested by modelling the C$^{17}$O
$J=2\to 1$ and $3\to 2$ lines observed with the JCMT, which, like the
submillimeter continuum, trace the total column density.  We assume a
CO/H$_2$ abundance of $2\times 10^{-4}$, as found for warm, dense
clouds by Lacy et~al.~\markcite{lac94}(1994). Chemical effects are
unlikely to change the CO abundance by more than a factor of 2 once
most of the gas-phase carbon is locked up in CO. Other carbon-bearing
molecules are at least $10^3$ times less abundant than CO. No solid CO
has been detected towards GL~2591 to a gas/solid CO ratio of $> 400$
(Mitchell et~al.~\markcite{mit90}1990, van Dishoeck
et~al.~\markcite{evd96b}1996). Estimates of the C$^0$ and C$^+$ column
densities are factors of $50-100$ lower than of that of CO (Choi
et~al.~\markcite{cho94}1994; C.~Wright, private communication).

The C$^{17}$O emission observed with the JCMT can be matched using a
range of values for $\alpha$, with the implied gas masses within a
radius of $30,000$~AU ranging from $33$~\msol\ for $\alpha=1.5$ to
$47$~\msol\ if $\alpha=1.0$. Comparing these numbers to the dust
masses from Table~\ref{phot_obs.tab}, it is seen that only the dust
opacities for coagulated grains (OH5 and OH2) yield the standard
gas-to-dust mass ratio of $100$. This is strong evidence that grain
coagulation occurs in circumstellar envelopes. The corresponding
density at $30,000$~AU is $1.5\times 10^4$~\ccm\ for
$\alpha=1.5$ and $4.5\times 10^4$~\ccm\ if $\alpha=1.0$. Although the
corresponding total masses are lower limits because the
power law may extend further out, the ratio of the dust and gas masses
is independent of choice of outer radius. Doubling the radius of the
model to $60,000$~AU, for instance, gives the same strength for the
C$^{17}$O lines within $0.5$~K, or $20$\%, while the dust emission is
even less affected (\S~\ref{s:dusto}). The optical depths of the
C$^{17}$O lines in the central pencil beam of the model vary from
$0.06-0.12$ for $\alpha=1.5$ to $0.09-0.15$ if $\alpha=1.0$.

Calculations were also performed for a source distance of 2~kpc, with
the model dimensions doubled, a luminosity of $8 \times 10^4$~\lsol,
and an effective temperature of 37,500~K. The radius where the dust
temperature has a certain value is found to double as well, so that
the temperature structure is constant in terms of projected size (in
arcseconds), but not in linear size (in AU). To match the observed
submillimeter continuum flux densities and C$^{17}$O line fluxes, the
required dust and \htwo\ masses are four times those for a
distance of 1~kpc. 

\subsection{Density Distribution}
\label{s:alpha}

The critical densities of the C$^{17}$O lines are such that they are
useful to constrain the total column density, but not to discriminate
between values of $\alpha$. One of the most suitable molecules to do
this is CS, which, because of the small spacing of its energy levels
and large dipole moment, samples a large range in critical densities
within the observable frequency range. Larger values of $\alpha$ imply
that more material is at high gas densities, increasing the ratio of
high-$J$ to low-$J$ emission. Further, the $J=5\to 4$~and $J=7\to
6$~lines of CS have been observed in several beams, constraining
$\alpha$ even better. The same lines are systematically brighter in
smaller beams, which is direct evidence for an inward increase of the
density.  Indeed, our data and those of Carr
et~al.~\markcite{car95}(1995) rule out a constant density model. The
H$_2$CO lines all have critical densities of $10^5-10^6$~cm$^{-3}$,
and are primarily sensitive to the temperature structure (e.g., Mangum
\& Wootten \markcite{man93}1993, van Dishoeck
et~al.~\markcite{evd93}1993). Specific \hhco\ line ratios can
therefore act as a check on the assumption that the kinetic
temperature equals the dust temperature. The observed HCO$^+$ and HCN
lines also have critical densities in the $10^5 - 10^7$ cm$^{-3}$
range and can be used as checks on the derived density structure. For
each combination of $\alpha$ and $n_0$, we first calculated the
temperature structure using the OH5 dust opacities, and then ran Monte
Carlo models for the line emission, tuning the molecular abundances to
minimize the difference between observed to synthetic line fluxes. The
quality of the fit was measured using the quantity $\chi^2$ as defined
by Zhou et al.  \markcite{zho91}(1991). The uncertainty in the data
was taken to be $30$\% for all lines except those in the $490$~GHz
window, where calibration is difficult, and those obtained in beams
$>20''$~FWHM with the CSO and NRAO telescopes, which have larger
pointing errors and may suffer from beam dilution. These latter lines
have an estimated accuracy of $50$\%.

It was found that \chisq\ has a minimum for $\alpha=1.0-1.5$,
especially for CS and \cs.  With $\alpha<1.0$, very little material is
at high densities, giving insufficient excitation of the high-$J$
lines. These are well matched for $\alpha \gtsim 1.5$, but now the
low-$J$ lines are a factor of $\sim 2$ weaker than observed.
Intermediate values of $\alpha$ provide the observed ratio of low-$J$
to high-$J$ emission and hence the lowest \chisq. The best-fitting
model has $n=3.5\times 10^{4} (r/30,000 {\rm AU})^{-1.25}$~cm$^{-3}$,
and an \htwo\ mass of 42~\msol. In the case of \hhco, the minimum of
\chisq\ is not as pronounced, which confirms that the temperature of
the molecular gas is close to that of dust, and reflects our finding
that the dependence of the dust temperature profile on density
structure is small (Figure~\ref{sed.fig}). The \chisq\ for HCN does
not increase for $\alpha > 1.5$, suggesting that this molecule is
confined to the inner parts of the source, where the density is high.
With only 3 observed lines, \hcop\ is not a sensitive probe of
$\alpha$.

The best-fitting abundances are $5\times 10^{-9}$ for CS, $2\times
10^{-9}$ for \hhco, $2\times 10^{-8}$ for HCN and $1\times 10^{-8}$
for HCO$^+$. The calibration of the data would give abundances to
$20$\% precision, but the major source of error is the CO abundance,
leading to an absolute uncertainty of a factor of 2.  The abundances
are an order of magnitude larger than those derived by Carr
et~al.~\markcite{car95}(1995), because the \htwo\ column density of
our model is lower.  The abundance ratios relative to CO are
unchanged, however, so that their discussion and conclusions are not
affected. The Carr et~al. density structure was not constrained with a
column density tracer, leading to a CO abundance of only $2\times
10^{-5}$. Since no significant depletion of CO in ice mantles is
observed toward GL 2591, this solution now appears less realistic.  In
addition, the densities in the models presented here are lower because
the temperatures are higher. While Carr et al. used an $r^{-0.4}$
temperature structure, the actual temperature found from our dust
modelling is higher in the inner region. We conclude that a
calculation of the temperature structure is a prerequisite for a
determination of the density structure and chemical composition of
high mass cores.

The value of $\alpha$ that fits the CS data best, $\alpha = 1.25$, is
not predicted by theoretical considerations, but it may result from
the averaging out of radial substructure. In very young objects, the
central part of the cloud should be undergoing infall (for which
$\alpha=1.5$), while further out, the original density profile is
maintained. In this interpretation, the data require that the density
law of the pre-collapse state is flatter than $r^{-1.5}$, and thus
significantly flatter than the $r^{-2}$ law expected for thermally
supported clouds.  Collapse in an initially logotropic sphere
(McLaughlin \& Pudritz \markcite{mcl97}1997), with $\alpha \sim 1$
before collapse, might reproduce the data. An average value of
$\alpha=1.25$ may also arise in more evolved stages, where even the
outer regions are all infalling ($\alpha=1.5$), while at the center,
heating and winds from the star push material out, causing an overall
flattening of the density structure (i.e., decrease the average
$\alpha$).

\subsection{Comparison to Infrared Data and Alternative Models}
\label{s:ch}

Another test of the power law model is a comparison to infrared
absorption line observations, which probe scales down to the point
where the observed $4.7$~\mic\ continuum forms. The high brightness of
this continuum is likely due to deviations from a spherical shape as
discussed in \S~\ref{s:2dm}.  Lada et al. (1984) fit the emission at
$\lambda \leq 13$ \mic\ with a shell of angular size 0\farcs06, and
speckle observations at 2.2 \mic\ give a size $<$0\farcs02 (Howell,
McCarthy \& Low \markcite{how81}1981).  These numbers are quite
uncertain, so we adopt a size of 0\farcs1, or 100~AU, for the minimum
radius probed by the infrared absorption, which is much smaller than
can be probed with the radio data.

Plotted in Figure~\ref{rot-synth.fig} is the column density implied by
our best-fitting spherical power law model in a pencil beam from cloud
center to edge in the \co\ rotational levels up to $J=25$ and the HCN
levels up to $J=11$, divided by their statistical weights. Superposed
are infrared observations for the \vlsr$=-5.7$~\kms\ component of \co\ 
from Mitchell et~al.~\markcite{mit89}(1989). The higher dispersion
data presented in this work are not shown since we have only a few
lines and they agree with the equivalent widths of Mitchell et al.
(1989).  The model points have been scaled to a total \co\ column
density of $2.1\times 10^{17}$~\scm, the value derived by Mitchell et
al. from a two-temperature fit to these data. This scaling accounts
for the difference in line width of a factor of 5 between the
rovibrational and the rotational lines.  The observed populations of
energy levels $\sim 1000$~K above ground are seen to be reproduced to
a factor of $\approx 3$, even though the highest temperature in the
model is only $350$~K, which illustrates that temperatures determined
from rotation diagrams do not directly reflect the kinetic
temperature. The observed column density in high-$J$ levels is even
better matched with $\alpha=1.5$, but this model falls short of the
low-$J$ population by an order of magnitude. The opposite is true for
$\alpha=1.0$; the conclusion is that both millimeter and infrared
observations on average support the value $\alpha \approx 1.25$.  The
factor of 3 discrepancy between data and model is somewhat larger than
the scatter in the data, and the two-temperature model by Mitchell et
al.~\markcite{mit89}(1989) provides a better fit. We interpret this
fact as evidence for a nonspherical geometry, which we discuss in more
detail in the next section.  In the case of CO, unlike HCN, infrared
pumping is unlikely to contribute to the excitation, since the
radiation field at the lowest-frequency vibrational band of CO at
$4.7$~\mic\ is weak (Fig.~\ref{sed.fig}).

To see if core-halo models can also match the rotational emission line
data, models have been run which employ the two-component structure
derived by Mitchell et~al.  (\markcite{mit89}1989,
\markcite{mit90}1990) from the \co\ data shown in
Figure~\ref{rot-synth.fig}. The halo and the core have temperatures of
40~K and 1000~K and CO column densities of $7.2 \times 10^{18}$~\scm\ 
and $5.6 \times 10^{18}$~\scm, respectively. The requirement that the
excitation is thermalized leads to typical sizes of the core of $3''$
and of the halo of $\le 60''$. Such models are unable to match the
observed strengths of the rotational lines, unless the abundances in
the core are higher than in the halo. As an example, the CS and
C$^{34}$S line profiles can be fitted for a core with diameter $3''$,
$n$(H$_2$)=$2.5\times 10^6$ cm$^{-3}$ and a CS abundance of $1\times
10^{-8}$, and a halo with diameter $48''$ (0.23~pc),
$n$(H$_2$)=$2\times 10^5$ cm$^{-3}$ and a CS abundance of $3\times
10^{-9}$. These models illustrate the effect of different physical
structures on the inferred abundances. Further observational and
theoretical research on the physical processes leading to such a
core-halo structure is needed.  Spectra at positions away from the
center should be able to distinguish between power laws and core-halo
models, but currently available spectra do not clearly rule out
core-halo models.

\subsection{Two-dimensional Models}
\label{s:2dm}

Although the model developed in \S\S~\ref{s:dusto}-~\ref{s:alpha}
matches the molecular line emission and the photometry in the
mid-infrared to submillimeter range well, several other observations
of GL~2591 remain unexplained. First, all dust models except that of
Mathis et al. \markcite{mat90}(1990) fail to reproduce the
near-infrared ($2.3-10$~\mic) part of the spectrum.  In low-mass
objects, bright near-infrared emission is sometimes ascribed to
thermal dust emission from a hot circumstellar disk (e.g., Adams, Lada
\& Shu \markcite{als87}1987), and sometimes to scattered light in the
case of a disk seen edge-on (e.g., Burns et~al.
\markcite{bur89}1989). For less embedded massive stars, it has been
attributed to dust inside an ionized region $10-20$~AU from the star,
where resonantly scattered Ly$\alpha$ photons heat the dust to $\sim
1000$~K (Wright \markcite{wri73}1973; Osterbrock
\markcite{ost91}1991).  However, all these explanations require a low
opacity of the spherical envelope, since otherwise the hot region will
be obscured by cold foreground dust (Butner
et~al.~\markcite{but91}1991, \markcite{but94}1994).  For instance, at
$3.6$~\mic, the models presented in Figure~\ref{sed.fig} have optical
depths of $1.5-15$ for $\alpha=1.0$ and $3.5-36$ for $\alpha=1.5$. It
is more likely that the failure of the other models indicates a
deviation from spherical symmetry which provides a low-opacity pathway
along our line of sight. The optical depth at $3.6$~\mic\ of the
Mathis et al. model is a factor of~3 lower than that of the OH2 model.
Imaging of GL~2591 at $2.2$~\mic\ (Tamura et al.
\markcite{tam91}1991) shows, in addition to the bright point source, a
loop of emission extending $5''$ to the West. This feature is
interpreted as a limb-brightened cavity cleared by the outflow. The
loop, also seen in VLA observations of \ammo\ by Torrelles et~al.
\markcite{tor89}(1989), is indeed coincident with the blue outflow
lobe (see Figure~\ref{jcmt_maps.fig}), i.e., the one directed towards
us.

Second, although the $\alpha=1.25$ model reproduces the total fluxes
of all our observed rotational lines, the synthetic line profiles of
CS, HCN, HCO$^+$ and $^{13}$CO are self-absorbed, contrary to
observation.  This same problem was encountered by Little
et~al.~\markcite{lit94}(1994) in their study of G34.3, but their
solution, artificially lowering the $^{12}$C/$^{13}$C ratio to
$\approx 20$, is not satisfactory.  Geometric effects are more likely
to play a role, since they affect only the more opaque transitions.
The deviation from spherical symmetry can either be global, e.g. in a
cavity formed by the stellar wind, or local, as unresolved density
variations, so that CS traces high-density ``clumps'' embedded in a
lower-density medium traced by CO. Alternatively, radial variations in
the molecular abundances can fix up the line shape. It is not the goal
of this paper to distinguish between these effects, which to first
order do the same, namely decrease the line opacity. However,
large-scale geometry must play some role as demonstrated by the
near-infrared emission from the dust, which is insensitive to both
chemistry and clumping.

A third, albeit weaker, piece of evidence for a nonspherical geometry
is provided by the infrared recombination lines of \ion{H}{1}. Tamura
\& Yamashita \markcite{tam92}(1992) detected Br$\gamma$ emission at
the infrared continuum position, with a flux of $8.2\times
10^{-17}$~W~m$^{-2}$ in a $(2.9\times 4.3)''$ beam, a factor of $5-10$
weaker than in similar objects (Bunn et al. \markcite{bun95}1995). The
Br$\gamma$ emission was found to be extended well beyond the radio
source: integrated over a $(10\times 16)''$ area, the Br$\gamma$ flux
is $2.2\times 10^{-16}$~W~m$^{-2}$. The Br$\alpha$ flux is $<2.3\times
10^{-15}$~W~m$^{-2}$ in an $11''$ beam (Simon et al.
\markcite{sim79}1979), consistent with the Br$\gamma$ to Br$\alpha$
ratio of $0.35$ for standard ``case B'' recombination, so that
scattering is probably unimportant at $2.16$~\mic. Using the fact that
the SW continuum source (Figure~1) is an optically thin \ion{H}{2}
region, Tamura \& Yamashita set a lower limit to the extinction
towards the SW source of $A_V>40$~mag, implying that the SW source is
located behind the GL~2591 molecular cloud. The extinction towards the
infrared source is not as straightforward to calculate since the
nature of the extended emission is not clear (Tamura \& Yamashita
\markcite{tam92}1992). In addition, the relation between radio
continuum and near-infrared \ion{H}{1} line emission is complex when
the realistic case of a non-spherical wind with partial ionization is
considered (Natta \& Giovanardi \markcite{nat93}1993). The ionized
emission from massive young stars with a weak radio continuum such as
GL~2591 warrants further study, but is outside the scope of this
paper.

The effect of a low-density cone around the central line of sight was
approximated by calculating the molecular excitation in the best-fit
spherically symmetric model, but with the density set to zero in a
cone of opening half-angle $\theta$ around the central line of sight.
The outflow axis was assumed to extend all the way from the inner to
the outer radius of the model, so that the model cannot be expected to
fit the data in detail, because cold gas is seen in absorption towards
this source. Synthetic spectra are obtained by performing radiative
transfer integrations through this two-dimensional model followed by
convolution to the appropriate telescope beam.  The results are
compared to the spherical case in Figure~\ref{mc-cs.fig} for a value
of $\theta = 30^\circ$, which gave somewhat better fits to the
observations than $\theta=15^\circ$, $45^\circ$ or $60^\circ$. We
conclude that there is ample evidence that the circumstellar material
is not spherically distributed and that a low-opacity pathway close to
our line of sight has important effects on the source's appearance.

\section{Modeling the interferometer data}
\label{s:o.models}

Using the physical structure of the gas and dust in the extended
envelope derived in the previous section, we will now investigate if
the best spherical model can also reproduce the interferometer line and
continuum data. We do not attempt to calculate the emission from a
two-dimensional model since this requires specifying a source function
(viz. temperature and density) in addition to an optical depth. This
introduces additional free parameters which cannot be independently
constrained. 

\subsection{Continuum emission}
\label{s:c.models}

\begin{figure}[h]

\centerline{\hbox{
\psfig{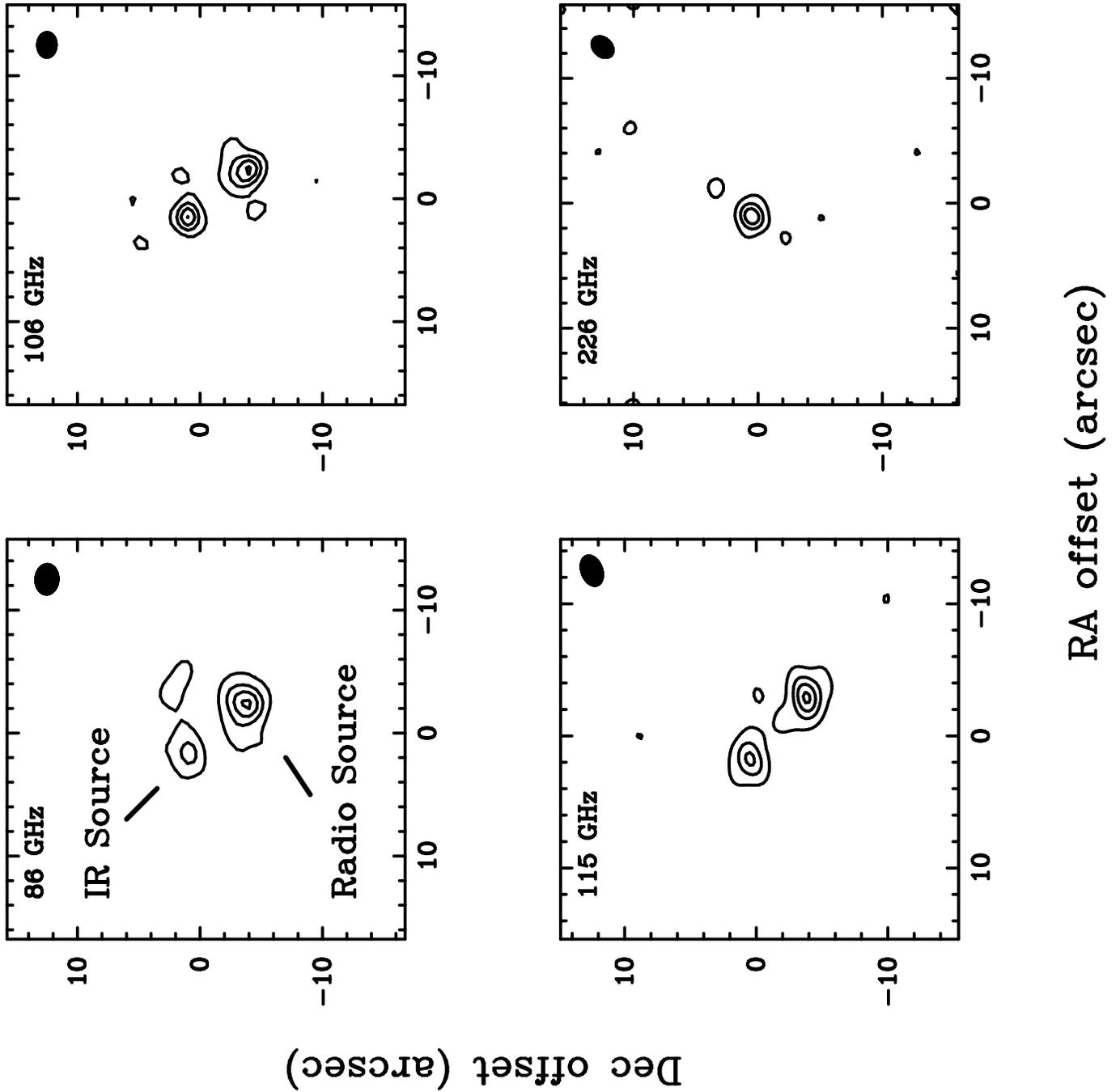}
}}
\noindent
  
  \caption{OVRO continuum maps of GL~2591. The beam FWHM is
  indicated in the top left corner. Contour levels are 6 to 42 by 12
  mJy~beam$^{-1}$ at 86~GHz, 5 to 35 by 10 mJy~beam$^{-1}$ at 106~GHz,
  10 to 50 by 20 mJy~beam$^{-1}$ at 115~GHz and 20 to 140 by 40
  mJy~beam$^{-1}$ at 226~GHz. The map center is at R.A. 20$^h$ 27$^m$
  35$^s$.8, Dec +40$^\circ$ 01$'$ 14$''$ (B1950).}
  \label{c_obs.fig}
\end{figure}

\begin{figure}[h]
  
\centerline{\hbox{
\psfig{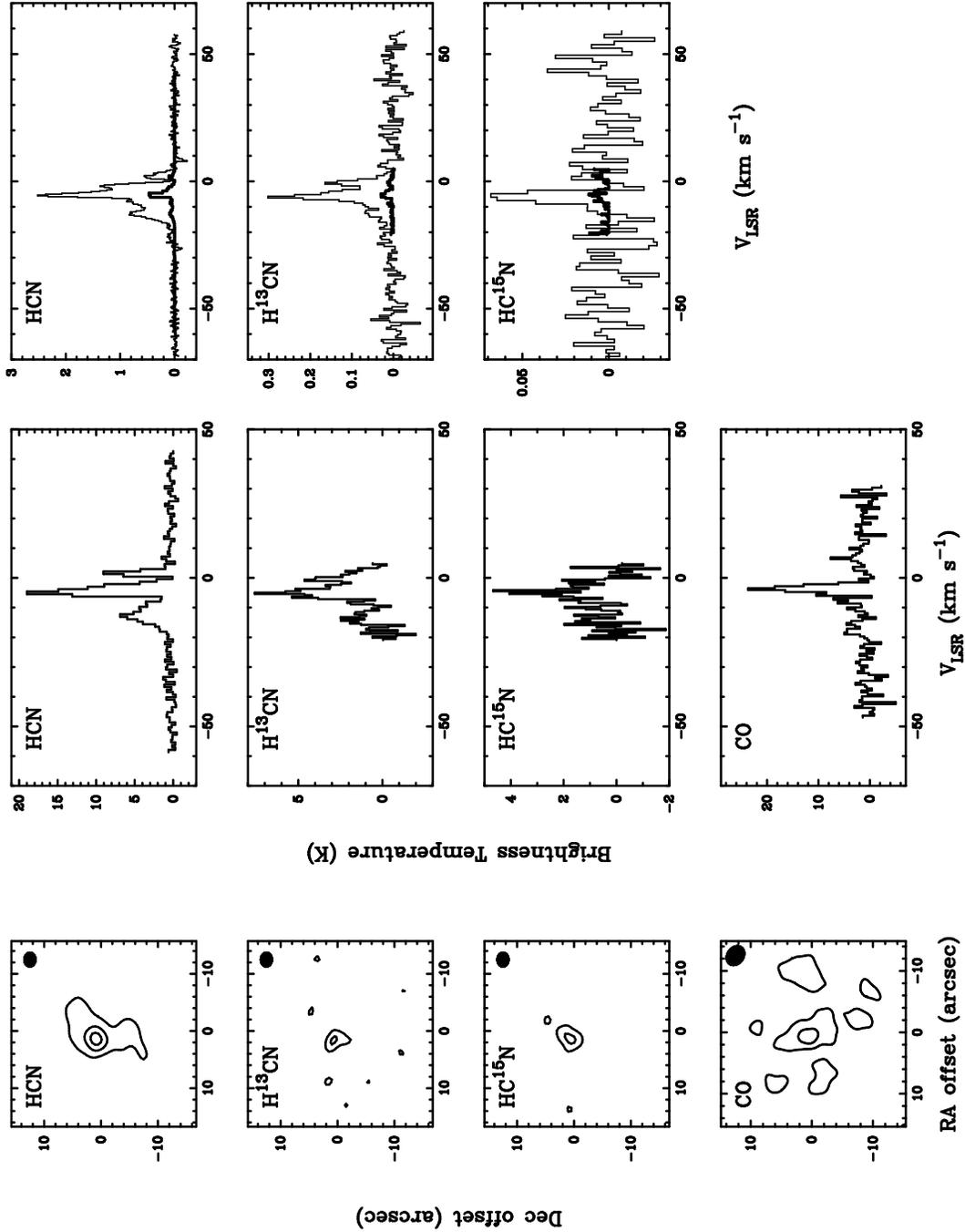}
}}

  \caption{Integrated intensity maps {\it (left
    panels)} and spectra {\it (middle panels)\/} of molecular lines
  observed with OVRO in a $3''.5$~beam. The map center is the same as
  in Figure~1. Lowest contour and contour interval is
  $130$~mJy~beam$^{-1}$ (HCN), $110$~mJy~beam$^{-1}$ (H$^{13}$CN),
  $80$~mJy~beam$^{-1}$ (HC$^{15}$N), $200$~mJy~beam$^{-1}$ (CO).  The
  right panels show spectra observed with the NRAO~12m telescope {\it
    (thin lines)\/} and with OVRO {\it (thick lines)}, both in a
  $63''$ beam.}
  \label{l_obs.fig}
\end{figure}

Unlike the case for single-dish observations, comparison of models
with interferometer data cannot proceed by calculating the total
emission within the beam. The measured flux densities in
Table~\ref{posflux.tab} are based on a deconvolution with the CLEAN
algorithm, which aims to find point sources. In observations of
extended sources at high frequencies, a large fraction of the emission
is on angular scales larger than that corresponding to the shortest
antenna spacing. For resolved objects, the models must be sampled and
deconvolved in a manner identical to the interferometer data, but
since GL~2591 appears circularly symmetric in our maps
(Figures~\ref{c_obs.fig} and~\ref{l_obs.fig}), a simple
one-dimensional comparison in the Fourier or $uv$ plane is sufficient.

Figure~\ref{cvis.fig} shows the calculated visibility function of the
$\alpha=1.0$~and $\alpha=1.5$~dust models, together with the
observations.  To obtain the visibility function from the data at
$86-115$~GHz, we subtracted the SW~source from the $uv$ data, shifted
the phase center to the infrared source, and binned the data in annuli
about the source. At $226$~GHz, the SW~source was not subtracted. The
error bars in the plot reflect only the 1$\sigma$ spread in the data
points. The bins at the shortest and longest baselines have
considerably more uncertainty because these regions of the $uv$ plane
are the most sparsely sampled by the interferometer. Note that the
high flux observed on the shortest baselines does not show up in the
images in Figure~\ref{c_obs.fig}, illustrating the limitations of the
deconvolution algorithm.

From Figure~\ref{cvis.fig}, the model is seen to reproduce the
visibility data only on the shortest spacings. On baselines $\gtsim
15$~\klm, the observed $86-115$~GHz flux lies a factor of $3-4$ above
the model value. If the emission is resolved at the largest spacing,
the radius of the compact source is $\approx 1000$~AU, in which case
the temperature of an opaque source equals the brightness temperature
of $\approx 1$~K. However, such a low temperature at this radius is
physically untenable because the warmer envelope fills most of the sky
as seen from the source and acts as an oven. Hence emission on this
scale must be optically thin, and the compact source must be at least as
warm as the ambient envelope. The 86 and 106 GHz data are actually
consistent with an unresolved source.  In this case, we derive a lower
limit to the source radius by following the curve $T_B(r/1000\ {\rm
  AU})^{-2}$ to where it intersects the temperature curve calculated
for the envelope.  This leads to an estimated radius of $\ltsim 30$~AU
and a temperature of $\gtsim 1000$~K for an opaque source.
Alternatively, the compact emission may be due to a steepening of the
density gradient for radii $<1000$~AU.  The $226$~GHz data are
uncertain, but appear well matched by the model power law envelope,
which ``outshines'' any possible compact source.  Higher-resolution
interferometry in the $100$~GHz band is needed to put additional
constraints on the size and orientation of this compact component.

\subsection{HCN line emission}
\label{s:l.models}

Infrared observations of GL~2591 by Carr et al.\ 
\markcite{car95}(1995) and by Lahuis \& van Dishoeck
\markcite{lah97}(1997) imply a $\gtsim 100\times$ higher abundance of
HCN than found in \S~\ref{s:alpha}. The HCN column density of
$1.5\times 10^{17}$ cm$^{-2}$ and excitation temperature $T_{\rm
  rot}\approx 1000$~K by Lahuis \& van Dishoeck should give rise to
strong rotational emission.  Do we see any emission from the inner
region reflected in the interferometer data?

The line emission from the envelope seen by OVRO was modelled using
the same approach as for the continuum, motivated by the structureless
appearance of the maps in Figure~\ref{l_obs.fig}. After solving for
the molecular excitation, maps of the sky brightness are constructed
at several velocities. These are Fourier transformed, and the result
is binned in annuli around the center.  The results are presented in
Figure~\ref{mc-ovro.fig}, which compares the observed and modelled
$J=1\to 0$ emission of all three HCN isotopes and of C$^{17}$O. To
avoid confusion with the broad, blueshifted velocity component, as
well as with the hyperfine components, the observed emission has been
integrated over just $1$~\kms, centered on \vlsr$=-5.7$~\kms.  For HCN
and H$^{13}$CN, the model results have been divided by 2 to account
for the hyperfine components. The poor match on short spacings is
probably due to inadequate sampling, so that the error bar was
underestimated.

The OVRO observations of HCN and isotopes show no excess at large
spacings over the predictions of the power law model, indicating that
the HCN abundance does not increase appreciably on radii down to
$1500$~AU, corresponding to the last data point at $80$~\klm.  To {\it
  avoid} the interferometer seeing the extra rotational emission from
the hot optically thick HCN observed with ISO, this region must be
beam diluted by at least a factor of 100 in the OVRO beam of 3\farcs5.
Hence, the source radius must be smaller than $175$~AU.

The temperature at $r=1500$~AU exceeds $120$~K (Fig.~\ref{sed.fig}).
Therefore, the source of HCN seen on smaller scales by ISO cannot
simply be the evaporation of icy grain mantles, since the most
refractory (and most abundant) interstellar ice, solid \water,
evaporates in $\sim 10$~yr when $T \gtsim 100$~K (Sandford \&
Allamandola \markcite{san93}1993). This is consistent with the fact
that solid HCN has not been detected down to levels of 3\% relative to
water ice, or $3\times 10^{-6}$ relative to \htwo\ (W.~Schutte,
private communication).  The temperature at $175$~AU from the star
exceeds 300 K, strongly suggesting that high-temperature chemistry is
producing the observed HCN enhancement.  One possibility is that most
of the oxygen is driven into \water\ at temperatures $>200$~K
(Charnley \markcite{cha97a}1997), leaving little atomic~O to destroy
HCN. The hot HCN seen with ISO is not the inner peak of the power-law
envelope because its rotation diagram is nearly flat (indicating $T
\sim 1000$~K), whereas the synthetic rotation diagram from the power
law model declines substantially over the observed range of lower
state energy (Fig.~\ref{rot-synth.fig}, bottom panel), indicating a
column-integrated excitation temperature of only $\sim 100$~K.  Thus,
the hot HCN must arise in a separate, HCN-enriched component, which is
probably very hot and dense.  The models presented in
Figure~\ref{rot-synth.fig} do not include infrared pumping, however,
while our detection of rotational lines of vibrationally excited HCN
suggests that this effect may contribute to the excitation.  Observations
at sub-arcsecond resolution of high-excitation lines are needed to
image this chemically active region.

\section{Conclusions}
\label{s:concl}

In this paper, the circumstellar environment of the massive young star
GL~2591 ($L = 2 \times 10^4$~\lsol, $D=1$~kpc) has been studied
through a variety of infrared, single-dish submillimeter, and
interferometric observations. The data are modelled using a two-step
method. The temperature structure is first calculated with a dust
radiative transfer code solving for the heating and cooling balance of
the grains as a function of distance to the star. The density
structure is then obtained with a Monte Carlo calculation of the
excitation of molecular emission lines, using the temperature
distribution calculated by the dust code.  Subsequently, this model
for the envelope on 30,000~AU scales is compared with the
interferometric and infrared spectroscopic data, which trace scales
from 100 to 30,000~AU.  Our main conclusions are written down below,
and sketched in Figure~\ref{cartoon.fig}.

1. The C$^{17}$O $J=2\to 1$ and $J=3\to2$ lines observed with the JCMT
indicate $A_V\approx 100$~mag, or a circumstellar mass of $\approx
42$~\msol\ contained within a volume of radius $30,000$~AU around the
star.

2. The radiative transfer calculations for the dust indicate that the
temperature is enhanced over the $T \propto r^{-0.4}$ relation valid
for optically thin dust inside a radius of $2000$~AU.  The dust mass,
obtained by fitting the far-infrared and submillimeter data, depends
strongly on choice of submillimeter dust opacity.  Consistency with
the gas mass derived from C$^{17}$O requires a high value for the
opacity; agreement to better than $50$\% is found adopting the
opacities by Ossenkopf \& Henning \markcite{oss94}(1994) for dust
grains which have coagulated and acquired ice mantles in the dense,
cold outer parts of the circumstellar envelope.

3. Single-dish observations of the CS and C$^{34}$S $J=2\to 1$ through
$10\to 9$ lines can be modelled successfully with a power-law
density structure $n=n_0 r^{-\alpha}$. The best match to the
\vlsr$=-5.7$~\kms\ lines is found for $\alpha=1.25\pm 0.25$ and
$n_0=(3.5\pm 1) \times 10^4$~\ccm\ at a radius of $30,000$~AU.  The
gas temperature as traced by emission lines of \hhco\ is found to
follow the dust temperature closely. The derived value of $\alpha$ is
between the values expected for free-fall collapse and nonthermal
support, suggesting a combination of these dynamical states within the
single-dish beams.

4. At $\sim 1500$~AU from the star, the calculated dust temperature
exceeds $120$~K, so that the ice mantles are expected to have
evaporated. Models with the ice removed from the grains indeed give
much better fits to the strong mid-infrared emission observed than do
models with ice-coated grains. Evaporation of the ice is also required
by the gas/solid ratios of H$_2$O and CO observed in the infrared.

5. The spherical power law model does not give a good match to the
strong near-infrared emission, to the line profiles of \co, CS, HCN
and HCO$^+$, and to the infrared recombination lines of \ion{H}{1}.
This suggests a deviation from spherical symmetry leading to a
decrease in opacity along the central pencil beam by a factor of
$\approx 3$, such as a low-density cone evacuated by the molecular
outflow, which is directed almost towards us. Models with an opening
angle of the cone of $\approx 30^\circ$ reproduce the CS, HCN and
\hcop\ line profiles. It appears that the star is actively shaping its
local environment.

6. We confirm the triple velocity structure in the $4.7$~\mic\ lines
of CO reported by Mitchell et al. \markcite{mit89}(1989), but measure
velocities of $(-23\pm 3.2)$ and $(-5.6 \pm 1.0)$ for the discrete
\co\ components, and a terminal velocity of $(173\pm 3)$~\kms\ for the
high-velocity wind. This velocity suggests an origin in envelope gas
entrained by the ionized jet seen in Br$\gamma$ and [\ion{S}{2}].  Due
to the finite size of the background continuum, both the quiescent
envelope gas (at $-5.6$~\kms) and the entrainment layer (seen as the
wind feature) are visible in absorption. The \co\ absorption at
$-21.5$~\kms\ arises in the foreground outflow lobe, which gas also
produces the wings on the rotational emission lines of CO, CS, \hhco,
HCN and HCO$^+$. The \co\ excitation seen in infrared absorption as
well as the strong CO $J=6\to 5$ emission indicates a temperature of
the outflow lobe of $\sim 200$~K.

7. The OVRO continuum data, when compared to the envelope model
derived from the single-dish data, reveal a compact source of radius
$< 1000$~AU, brightness temperature of $\approx 1$~K and spectral
index $\approx 1.7 \pm 0.3$.  No constraints on the geometry of this
component are available.  One possibility is optically thick, thermal
emission from a $\sim 1000$~K, $30$~AU radius region of hot dust.
Alternatively, the compact emission may reflect a steepening of the
density gradient on radii $<2000$~AU.

8. ISO observations by Lahuis \& van Dishoeck \markcite{lah97}(1997)
indicate an HCN abundance of $\sim 10^{-6}$ and an excitation
temperature of $\approx 1000$~K. The OVRO HCN data are consistent with
the power law envelope model, implying that the HCN abundance remains
at the $10^{-8}$ level down to radii of $\sim 1500$~AU.  Therefore,
the enhancement of HCN does not start until well after temperatures of
$120$~K are reached, considerably above the evaporation temperature of
any interstellar ice. The enhancement seems to occur within $175$~AU
from the star where $T > 300$ K, suggesting that high-temperature
gas-phase chemistry is important. This region is further evidence for
the effect of the star on its environment.

\acknowledgements

The authors are grateful to Michiel Hogerheijde for useful discussions
and for his efforts on the Monte Carlo radiative transfer code, to
Mario van den Ancker and G\"oran Sandell for providing their
unpublished JCMT/UKT14 data and for reducing the SCUBA data, to Craig
Kulesa for communicating his Phoenix results, and to an anonymous
referee whose comments helped to improve this paper.  They also would
like to thank the staffs of the OVRO, CSO, JCMT, NRAO~12m and
NOAO~2.1m telescopes for their support, especially Remo Tilanus and
Fred Baas at the JCMT and Ken Hinkle, Dick Joyce, and Jeff Valenti at
the NOAO~2.1m.  Byron Mattingly carried out the NRAO 12m observations.

FvdT is grateful to the Leids Sterrewacht Fonds, the Stimuleringsfonds
Internationale Betrekkingen of the Netherlands Organisation for
Scientific Research (NWO), and the Leids Kerkhoven-Bosscha Fonds for
travel support.  Astrochemistry in Leiden is supported by grant
781--76--015 from the Netherlands Foundation for Research in
Astronomy. N.~J.~E. acknowledges support from NSF Grant AST-9317567
and the Randall Professorship. G.~A.~B. gratefully acknowledges
support provides by NASA grants NAG5-4383 and NAG5-3733.

\vfill
\eject

\begin{deluxetable}{llll}
\tablecaption{Parameters of OVRO observations. \label{ovro_log.tab}}

\tablehead{
 & 
\colhead{Setting 1} & 
\colhead{Setting 2} & 
\colhead{Setting 3}}

\startdata

Frequency (GHz) & 86 & 106 & 115/226 \nl
Observation date & 27sep95/16dec95 & 29sep95/02jan96/03jan96 & 01feb97/24mar97/01jun97 \nl
Configuration\tablenotemark{(a)} & LR/EQ & LR/EQ/EQ & LR/HR/LR \nl
Time on-source (min.) & 380/460 & 420/180/235 & 260/140/280 \nl

Wide mode: & & & \nl
\ \ Resolution (\kms) & 0.87 & 0.71 & 0.65 \nl
\ \ Coverage (\kms) & 112 & 90.5 & 83.4 \nl
\ \ Lines        & HCN $J=1\to 0$, & 
                   CH$_3$OH $3_1-4_0$ A$^+$, &
                   CO, C$^{17}$O \nl
                 & SO $J=2_2 - 1_1$ & SO$_2$ $10_{1,9}-10_{0,10}$ & $J=1\to 0$ \nl
Narrow mode: & & & \nl
\ \ Resolution (\kms) & 0.44 & 0.35 & 0.33 \nl
\ \ Coverage (\kms) & 27.9 & 22.6 & 20.9 \nl
\ \ Lines       & H$^{13}$CN, HC$^{15}$N &
                     CH$_3$OH $11_{-1}-10_{-2}$ E, &
                     CO, C$^{17}$O \nl
                   & $J=1\to 0$ & H$_2$CS $3_{12}-2_{11}$ & $J=1\to 0$ \nl
Continuum bandwidth & 0.5 GHz & 0.5 GHz & 1.0 GHz \nl
Longest baseline (\klm) & 70 & 84 & 95/90 \nl 
Synthesized beam FWHM: & & & \nl
\ \ Continuum & 
$(2.4 \times 1.8)''$, $-88^\circ$ &
$(2.0 \times 1.5)''$, $-88^\circ$ &
$(2.3 \times 1.5)''$, $-69^\circ$ / \nl
& & &                 
$(1.9 \times 1.4)''$, $-43^\circ$ \nl
\ \ Lines &           
$(3.7 \times 3.0)''$, $-73^\circ$ & \nodata &
$(3.4 \times 2.7)''$, $-54^\circ$ \nl
Map rms (mJy beam$^{-1}$) & 60 (CO, HCN) &
80-100 (H$^{13}$CN, HC$^{15}$N) & 1.7 (continuum) \nl
(all settings) & & & \nl

\enddata

\tablenotetext{(a)}{The LR configuration is compact, the HR is
  extended, and the ``equatorial'' (EQ) configuration has long
  North-South but shorter East-West baselines. }

\end{deluxetable}

\vfill
\eject

\vfill

\begin{deluxetable}{lllll}
\tablewidth{0pt}
\scriptsize
\tablecaption{Positions and Flux Densities of Radio Sources in GL~2591.\label{posflux.tab}}

\tablehead{
 & 
\colhead{RA (1950)} & 
\colhead{Dec (1950)}  & 
\colhead{Flux Density} & 
\colhead{Reference} \nl
 & 
\colhead{20$^h$27$^m$} & 
\colhead{+40$^\circ$01$'$} & 
\colhead{   (mJy)} & 
          }
\startdata
 & & & & \nl
\multicolumn{5}{c}{SW source} \nl
 & & & & \nl
 VLA 5~GHz      & 35$^s$.613 (0.007) & 10$''$.4 (0.1) & \ 79 (2.0) & Campbell (1984) \nl
 VLA 8.4 GHz    & 35$^s$.659 (0.007) & 10$''$.4 (0.1) & \ 82 (8.0) & Tofani et~al. (1995) \nl
 & & & & \nl     
 OVRO 87 GHz    & 35$^s$.63 (0.004) & 10$''$.4  (0.1) & \ 87 (1.4) & This work \nl
 OVRO 106 GHz   & 35$^s$.62 (0.004) & 10$''$.4  (0.1) & \ 71 (1.2) & This work \nl
 OVRO 115 GHz   & 35$^s$.63 (0.01)  & 10$''$.4  (0.1) & \ 93 (2.5) & This work \nl
 
 & & & & \nl
\multicolumn{5}{c}{NE source} \nl
 & & & & \nl
 VLA 5~GHz      & 35$^s$.963 (0.007) & 14$''$.8 (0.1) & \ \ 0.4 (0.1)  & Campbell (1984) \nl
 VLA 8.4~GHz    & 35$^s$.975 (0.007) & 14$''$.7 (0.1) & \ \ 0.5 (0.1) & Tofani et~al. (1995) \nl
 & & & & \nl                                                    
 OVRO 87 GHz    & 35$^s$.95  (0.004) & 14$''$.9 (0.1) & \ 29.5 (0.8) & This work \nl
 OVRO 106 GHz   & 35$^s$.93  (0.003) & 14$''$.9 (0.1) & \ 38.7 (0.7) & This work \nl
 OVRO 115 GHz   & 35$^s$.97  (0.009) & 14$''$.8 (0.1) & \ 52.9 (1.5) & This work \nl
 OVRO 226 GHz   & 35$^s$.88  (0.013) & 14$''$.7 (0.2) & 151  (4.5)\tablenotemark{a} & This work \nl
 & & & & \nl                                                                     
 2.2~\mic       & 36$^s$.00  (0.09)  & 15$''$ (1)  &   & Tamura et~al. (1991) \nl

\enddata
\tablenotetext{a}{Includes $\le 80$~mJy from the SW source (see text).}
\end{deluxetable}

\vfill
\eject

\begin{deluxetable}{llccccc} 
\tablewidth{0pt}
\scriptsize
\tablecaption{Gaussian fits to observed submillimeter emission lines \label{obs_submm.tab}}

\tablehead{
\colhead{Molecule} &
\colhead{Transition} &
\colhead{Frequency} &
\colhead{$\int T_{\rm MB} dV$} &
\colhead{$\Delta V$} &
\colhead{$T_{\rm mb}$} &
\colhead{Telescope / Date} \nl
 & &
\colhead{(MHz)} &
\colhead{(K \kms)} &
\colhead{(\kms)} &
\colhead{(K)} &
}

\startdata

CI         & $^3P_1-^3P_0$ & 492160.7 & 85.8  & 8.64 & 9.34 & JCMT Oct '95 \nl
           & $^3P_1-^3P_0$\tablenotemark{a} &  -5.4   & 23.2  & 3.08 & 7.09 & \nl
           &           &  -8.3   & 68.9 & 11.56 & 5.60 & \nl
$^{13}$CO  & 3-2 & 330588.1 &  224.1 & 5.15 & 40.86 &  JCMT May '95  \nl
           & 3-2\tablenotemark{a} & -5.5 & 136.7 & 3.33 & 31.38 &   \nl
           &           & -7.0 & 111.0 & 8.77 & 14.64 &   \nl
           & 6-5 & 661067.4 & 236.0 & 5.51 & 40.3 & CSO May '95 \nl
           & 6-5\tablenotemark{a} & -5.5 & 113.5 & 3.75 & 28.5 & \nl
           &                      & -6.5 & 140.3 & 8.81 & 15.0 & \nl
C$^{17}$O & 2-1 & 224714.4 & 9.6 & 4.14 & 2.17 &  JCMT May '95 \nl
            & 2-1\tablenotemark{a} & -5.7 & 4.4 & 2.75 & 1.50  &  \nl
            &           & -7.2 & 5.6 & 5.36 & 0.98  &  \nl
            & 3-2 & 337060.9   & 14.7 & 3.28 & 4.19 &  JCMT May '95 \nl
            & 3-2\tablenotemark{a} & -5.7 & 6.03 & 2.14 & 2.66 & \nl
            &           & -6.6 & 9.10 & 4.90 & 1.74 & \nl
CS       & 5-4 & 244935.6 & 25.2 & 3.62 & 6.55 & JCMT Oct '95 \nl
     & 5-4\tablenotemark{a} & -5.6    & 11.7 & 2.45 & 4.46 & \nl
     &           & -6.5    & 14.6 & 5.13 & 2.70 & \nl
           & 7-6 & 342883.0 & 22.8 & 3.32 & 6.47 & JCMT May '96 \nl
     & 7-6\tablenotemark{a} &  -5.3   & 10.1 & 2.25 & 4.22 &  \nl
     &           &  -6.1   & 13.9 & 4.70 & 2.77 &  \nl
C$^{34}$S & 5-4 & 241016.2 & 3.41 & 3.16 & 1.01 & JCMT May '96 \nl
            & 7-6 & 337396.7 & 1.86 & 2.73 & 0.64 & JCMT May '95 \nl
 CS (v=1)    & 7-6 & 340398.1 & - & - & $<$ 0.20 & JCMT May '95 \nl

\hhco\    & $3_{03}-2_{02}$ & 218222.2 & 9.84 & 3.65 & 2.54 & JCMT Oct '95 \nl
          & $3_{03}-2_{02}$ \tablenotemark{a} & -5.8 & 4.51 & 2.17 & 1.96 &  \nl
          &                 &       -6.5 & 6.29 & 6.20 & 0.95 &  \nl
          & $3_{22}-2_{21}$ & 218475.6 & 2.57 & 3.83 & 0.63 & JCMT Oct '95 \nl
         & $3_{12}-2_{11}$ & 225697.8 & 14.44 & 3.81 & 3.56 & JCMT May '95 \nl
         & $3_{12}-2_{11}$ \tablenotemark{a} & -5.6 & 8.57 & 2.77 & 2.90 &  \nl
       &                   &     -6.6 & 7.11 & 6.80 & 0.98 &  \nl
       & $5_{15}-4_{14}$ & 351768.7 & 12.86 & 3.58 & 3.38 & JCMT May '95 \nl 
       & $5_{05}-4_{04}$ & 362735.9 &  5.71 & 3.21 & 1.67 & JCMT May '95  \nl
     & $5_{24}-4_{23}$       & 363945.9 & 3.55 & 3.57 & 0.93 & JCMT May '95 \nl
     & $5_{42/41}-4_{41/40}$ & 364102.8 & 1.00 & 4.96 & 0.19 & JCMT May '95 \nl
     & $5_{33}-4_{32}$ & 364275.2 & 4.24 & 4.41 & 0.90 & JCMT May '95 \nl
     & $5_{32}-4_{31}$ & 364289.0 & 4.34 & 4.86 & 0.84 & JCMT May '95 \nl
     & $7_{17}-6_{16}$ & 491968.9 & 9.98 & 4.76 & 1.96 & JCMT Oct '95  \nl
H$_2^{13}$CO & $5_{05}-4_{04}$ & 353811.8 & - & - & $<$ 0.20 & JCMT May '96 \nl

\tablebreak

HCN            & 4-3 & 354505.5 & 45.7 & 4.60 & 9.33 & JCMT May '95  \nl
               & 4-3\tablenotemark{a} & -5.1 & 24.7 & 3.47 & 6.67 &   \nl
               &                      & -6.4 & 23.1 & 6.52 & 3.33 &   \nl
HCN ($\nu_2=1$)         & $4_1-3_1$ & 354460.4 & 2.47 & 5.38 & 0.43 & JCMT May '95  \nl
H$^{13}$CN      & 1-0\tablenotemark{b} & 86340.2 & 1.78 & 3.63 & 0.26 & KP12m Nov '95 \nl
                & 3-2 & 259011.8 & 5.51       & 4.19 & 1.23 & JCMT Oct '95  \nl
                & 4-3\tablenotemark{c} & 345339.8 & 8.64 & 4.94 & 1.64 & JCMT May '95  \nl
HC$^{15}$N      & 1-0  & 86055.0 & 0.31 & 3.76 & 0.08 & KP12m Nov '95 \nl
                & 3-2 & 258157.3 & 2.39 & 4.38 & 0.51 & JCMT Oct '95 \nl
                & 4-3 & 344200.3 & 3.10 & 4.50 & 0.65 & JCMT May '95 \nl
HCO$^+$        & 4-3 & 356734.3 & 91.8 & 4.56 & 19.0 & JCMT May '96 \nl
         & 4-3\tablenotemark{a} &  -5.4   & 55.8 & 3.45 & 15.2 & \nl
         &                      &  -6.7   & 44.2 & 8.22 & 5.03 & \nl
H$^{13}$CO$^+$ & 3-2 & 260255.5 & 6.18 & 3.17 & 1.83 & JCMT Oct '95 \nl 
               & 3-2\tablenotemark{a} & -5.5 & 5.45 & 2.66 & 1.93 & \nl
               &                      & -8.1 & 0.88 & 2.04 & 0.40 & \nl
               & 4-3 & 346998.5 & 5.45 & 2.45 & 2.08 & JCMT May '96 \nl
               & 4-3\tablenotemark{a} & -5.4 & 4.37 & 2.11 & 1.95 & \nl
               &                      & -7.1 & 1.43 & 3.62 & 0.37 & \nl

HC$^{18}$O$^+$ & 4-3 & 340633.0 & - & - & $<$ 0.15 & JCMT May '96 \nl

\enddata

\tablenotetext{a}{Two-component fit, velocities (\kms) given instead
  of frequency.}  \tablenotetext{b}{Line integral is the sum of the
  hyperfine components; peak brightness refers to main component.}
\tablenotetext{c}{Blend with the SO$_2$ $13_{2,12}-12_{1,11}$ line.}

\tablecomments{Uncertainties of the line flux and brightness are $30$\% for JCMT
observations, due to calibration; the less stable pointing of the CSO and NRAO leads
to an error estimate of $50$\%. Peak velocities and line widths are accurate to 
$\approx 0.1$~\kms.}

\end{deluxetable}

\vfill
\eject

\begin{deluxetable}{rrrrrr}
\tablewidth{0pt}
\scriptsize
\tablecaption{Measured Properties of the Infrared Lines. \label{irvel.tab}}

\tablehead{
\colhead{Line} & 
\colhead{Rest frequency} & 
\colhead{Observed frequency} & 
\colhead{LSR Velocity} & 
\colhead{Line FWHM} & 
\colhead{Equivalent Width\tablenotemark{c}} 
\nl
 &  
\colhead{\rcm\ }         &
\colhead{   \rcm\ }           & 
\colhead{\kms\ }   & 
\colhead{\kms\ } &
\colhead{$10^{-3}$ \rcm\ }     }

\startdata

$^{12}$CO $v=1\gets 0$ R(1) & 2150.856 & 2151.199 &\ -19.1 & \nodata & 1.732 \nl
                          &          & 2152.978\tablenotemark{a} & -267 & \nodata & -0- \nl
R(2)                      & 2154.596 & 2154.938 &\ -18.9 & \nodata & 1.470 \nl
                          &          & 2156.213 & -196 & \nodata & -0- \nl
R(3)                      & 2158.300 & 2158.645 &\ -19.2 & \nodata & 1.214 \nl
                          &          & 2159.931 & -198 & \nodata & -0- \nl

P(2)                      & 2135.546 & 2135.710 & -20.8 & \nodata & 2.70 \nl
P(3)                      & 2131.632 & 2131.818 & -23.9 & \nodata & 2.81 \nl
                          &          & 2133.024 & -194 & \nodata & -0- \nl

P(7)                      & 2115.629 & 2115.817 & -24.4 & \nodata & 2.408 \nl
                          &          & 2116.605\tablenotemark{a} & -136 & \nodata & -0- \nl
P(8)                      & 2111.543 & 2111.731 &\ -24.5 & \nodata & 2.120 \nl
                          &          & 2112.933 & -195.1 & \nodata & -0- \nl
 & & & & \nl

$^{12}$CO $v=2\gets 1$ R(8) & 2149.489 & 2149.688 & +1.0 & 10.2 & 5.25 \nl
R(9)                      & 2152.942 & 2153.344 & -9.3  & 19.9 &  16.25 \nl
                          &          & 2153.215 & -27.2 & 19.1 &   4.60 \nl

R(10)                     & 2156.358 & 2156.566 &  -1.8 & 19.2 &  28.59 \nl
R(11)                     & 2159.739 & 2160.036 & -12.5 &  9.7 &   4.25 \nl


\tablebreak

$^{13}$CO $v=1\gets 0$ R(3) & 2110.442 & 2110.507 &  -7.0 &  9.5 & 33.39 \nl
                            &          & 2110.628 & -24.2 & 22.7 & 68.90 \nl
R(4)\tablenotemark{a}       & 2113.953 & 2114.043 & -10.5 &  6.7 & 21.98 \nl
                            &          & 2114.162 & -27.4 & 22.0 & 57.66 \nl
R(5)\tablenotemark{b}       & 2117.431 & 2117.518 & -10.1 & 18.8 & 68.86 \nl 
                            &          & 2117.628 & -25.7 &  8.1 & 21.00 \nl

R(9)\tablenotemark{a}      & 2131.004 & 2131.084 &  -9.0 & 52.5 & 50.91 \nl
R(10)                       & 2134.313 & 2134.354 &  -3.5 & 23.2 & 14.99 \nl
                            &          & 2134.535 & -28.9 & 34.8 & 59.17 \nl
R(11)                       & 2137.588 & 2137.668\tablenotemark{a} & -9.0 & 6.0 & 3.53 \nl
                            &          & 2137.809 & -28.8 & 20.6 & 35.57 \nl

R(15)\tablenotemark{a}      & 2150.341 & 2150.612 &  -9.0 & 10.7 &  2.89 \nl
R(16)                       & 2153.443 & 2153.696 &  -6.5 & 20.1 &  8.05 \nl        
                            &          & 2153.822 & -24.0 & 24.5 & 11.69 \nl
R(17)\tablenotemark{a}      & 2156.509 & 2156.805 & -12.4 & 30.0 & 23.70 \nl
R(18)\tablenotemark{a}      & 2159.540 & 2159.799 &  -7.2 & 40.7 & 25.65 \nl

\enddata

\tablenotetext{a}{Contaminated by telluric absorption.}
\tablenotetext{b}{On the edge of the spectrum.}  
\tablenotetext{c}{For
  the $^{12}$CO lines, which are well-resolved, the maximum optical
  depth is given. For the line wings, the -0- symbol means that the
  velocity in column 4 is where the line reaches the continuum.}

\tablecomments{When measured through profile fitting, velocities and
  line widths are accurate to $\approx 1$~\kms; terminal velocities
  were estimated by eye and have an uncertainty of 3~\kms.  Equivalent
  widths are accurate to $\approx 30$\% for clean lines and to $50$\%
  for contaminated lines.}

\end{deluxetable}

\vfill
\eject

\vfill

\begin{deluxetable}{r r r c c c c}
\tablecaption{Observed and modelled submillimeter flux densities of GL~2591. \label{phot_obs.tab}}

\tablehead{
\colhead{Dust model} &
\colhead{$\alpha$} & 
\colhead{Dust Mass}  &
\colhead{$F_{20}$} &
\colhead{$F_{450}$} &
\colhead{$F_{800}$} &
\colhead{$F_{1100}$} \nl
           & 
        &
\colhead{(\msol)} &  
\colhead{(Jy)}   & 
\colhead{(Jy)}    &  
\colhead{(Jy)}   &  
\colhead{(Jy)}     
   }

\startdata

       LG  & $1.5$ & $  3.22 $ & $   0.22 $ & $77.9$ & $6.75$ & $2.62$  \nl 
       LG  & $1.0$ & $  4.46 $ & $   0.22 $ & $71.3$ & $6.20$ & $2.47$  \nl 
       DL  & $1.5$ & $ 13.34 $ & $  20.8  $ & $75.3$ & $7.30$ & $2.70$  \nl 
       DL  & $1.0$ & $ 18.47 $ & $  39.8  $ & $69.6$ & $6.77$ & $2.56$  \nl 
       MMP & $1.5$ & $  1.18 $ & $ 288.   $ & $77.8$ & $5.30$ & $1.89$  \nl 
       MMP & $1.0$ & $  1.68 $ & $ 345.   $ & $73.2$ & $5.01$ & $1.83$  \nl
       OH5 & $1.5$ & $  0.34 $ & $ 139.   $ & $66.9$ & $6.12$ & $2.62$  \nl 
       OH5 & $1.0$ & $  0.48 $ & $ 265.   $ & $65.3$ & $5.97$ & $2.62$  \nl 
       OH2 & $1.5$ & $  0.21 $ & $ 731.   $ & $61.6$ & $7.71$ & $3.57$  \nl
       OH2 & $1.0$ & $  0.33 $ & $ 912.   $ & $67.8$ & $8.47$ & $4.00$  \nl

 & & & & \nl
Observed:  & && $733. \pm 147.$ & $66.3\pm 3.1$& $8.2\pm 0.24$  & $2.93 \pm 0.12$    \nl

\enddata

\tablecomments{
The $450-1100$~\mic\ data are JCMT observations by G.~Sandell (private communication). 
Beam FWHM is $18''.5$~at $450$~\mic, $17''.5$~at $850$~\mic\ and $18''.5$~at 
$1100$~\mic. The $20$~\mic\ flux is from Aitken et~al.~(1988), obtained in a 
$4''.2$ FWHM beam with UKIRT. The grain models are labelled as follows: 
DL = Draine \& Lee (1984), LG = Li \& Greenberg (1997), MMP = Mathis, Mezger \& 
Panagia (1983), OH = Ossenkopf \& Henning (1994). The masses refer to a volume
of radius 30,000~AU.}

\end{deluxetable}

\vfill
\eject

\input psfig

\vfill \eject

\vfill \eject
\begin{figure}[h]
  
\centerline{\hbox{
\psfig{figure=fig3.ps,height=18cm,angle=0}
}}

  \caption{Maps of GL~2591 made at the JCMT, centered
  at the same position as Figures~1 and~2. The beam size is
  $7\farcs5$~FWHM at $450$~\mic\ and $14\farcs3$~at $850$~\mic\ and in
  \co, while the peak brightness is $34$~Jy~beam$^{-1}$ at $450$~\mic,
  $6.7$~Jy~beam$^{-1}$ at $850$~\mic\ and $144$~K~\kms\ in \co.  {\it
    Top panels:\/} Submillimeter continuum emission, mapped with SCUBA
  at two wavelengths. Lowest contour and contour interval are $0.4$
  and $0.8$ Jy beam$^{-1}$ at $850$~\mic\ and $3.0$ and $6.0$ Jy
  beam$^{-1}$ at $450$~\mic.  {\it Bottom panels:\/} Raster map of
  \co\ $J=3\to 2$ emission, obtained with receiver B3i.  The left
  panel shows the total intensity, integrated from -15 to +5 \kms,
  with contours every 20 K \kms, starting at 40 K \kms. The right
  panel shows the line wings, integrated from -9 to -15 \kms\ (solid
  contours), and from +5 to -3 \kms\ (dashed contours). Lowest contour
  and contour interval is 5 K~\kms\ for both velocity ranges.}
  \label{jcmt_maps.fig}
\end{figure}


\vfill \eject
\begin{figure}[h]
  
\centerline{\hbox{
\psfig{figure=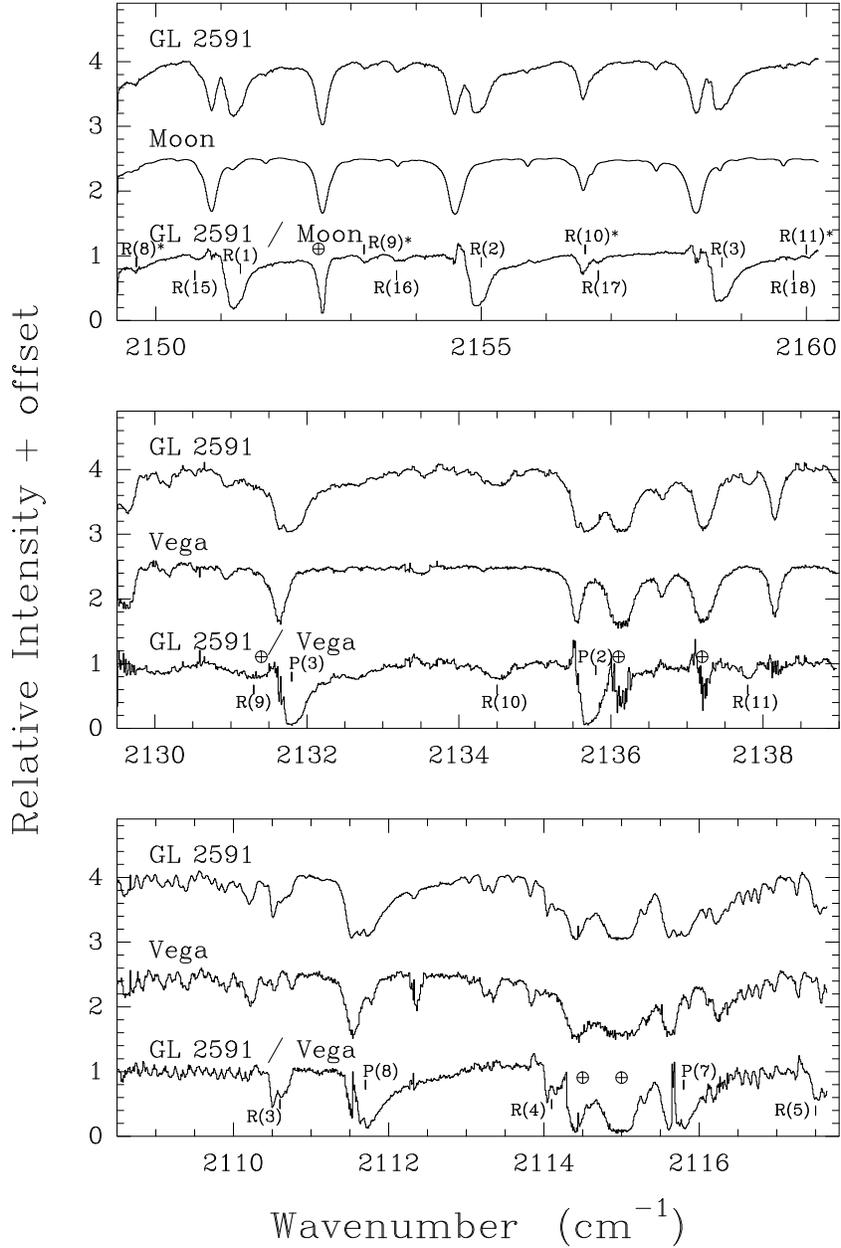,height=18cm,angle=0}
}}

  \caption{Spectra in three wavelength regions near
  $4.7$~\mic, observed with the Phoenix spectrometer. To eliminate
  telluric features, the source data are divided by the standard star
  data (Moon/Vega), scaled to the appropriate air mass. Lines of CO
  and \co\ in the $v=1\gets 0$ band are indicated above and below the
  spectrum, respectively. Lines in the $^{12}$CO $v=2\gets 1$ band are
  denoted with an asterisk.}
  \label{phxdata.fig}
\end{figure}


\vfill \eject
\begin{figure}[h]
  
\centerline{\hbox{
\psfig{figure=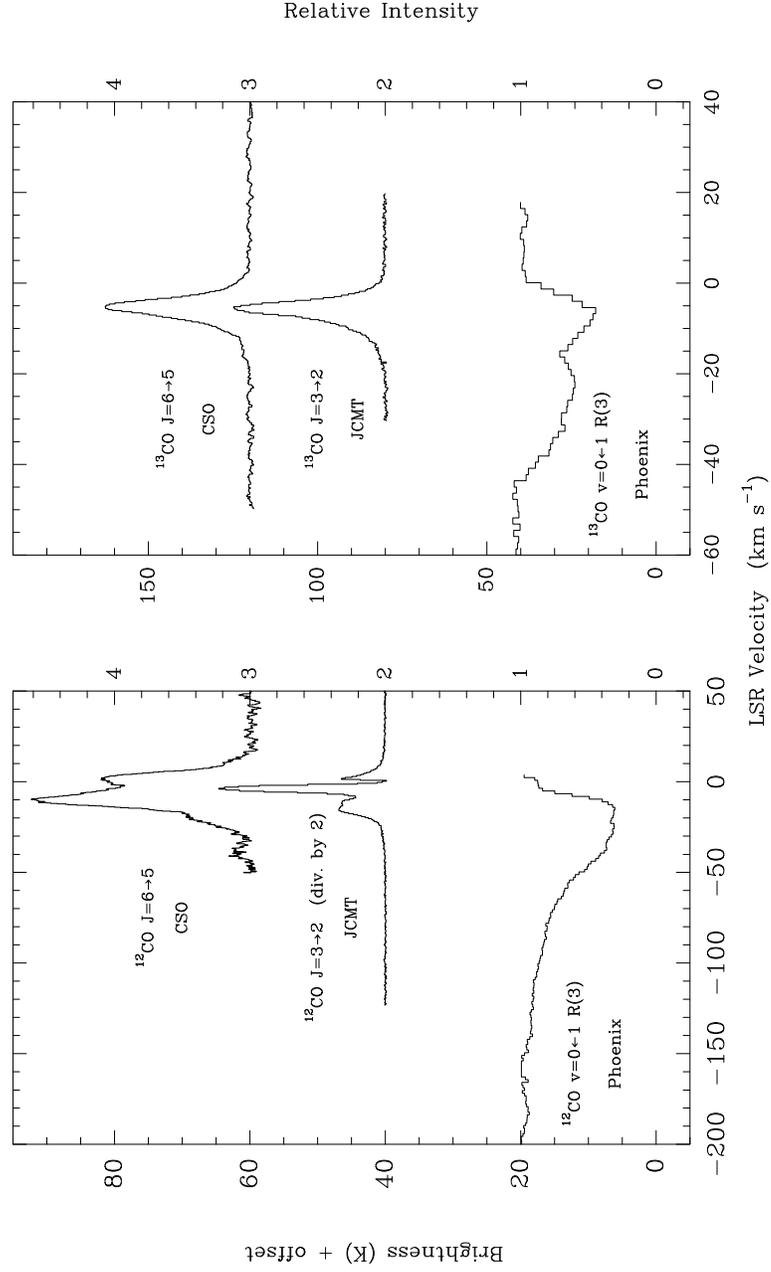,height=18cm,angle=0}
}}

  \caption{Submillimeter emission lines (in K) and
  infrared absorption lines (in relative units) of CO and \co,
  observed with the indicated telescopes.}
  \label{ir-submm.fig}
\end{figure}

\vfill \eject
\begin{figure}[h]
  
\centerline{\hbox{
\psfig{figure=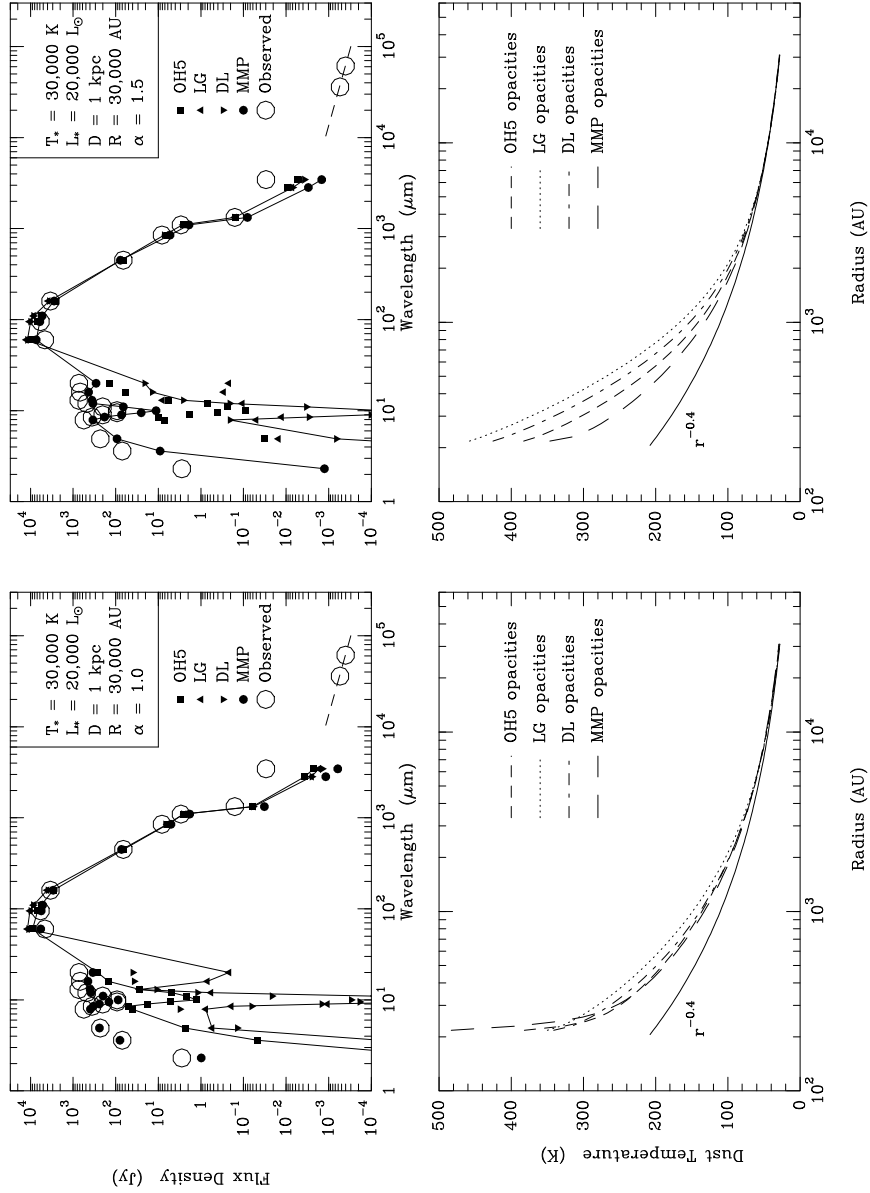,height=18cm,angle=0}
}}

  \caption{Observed continuum emission from
  GL~2591 ({\it large open circles\/}) with model fits superposed. In
  the left panels, the density drops as $r^{-1.0}$ with distance from
  the star $r$, while in the right panels, it drops as $r^{-1.5}$.
  {\it Bottom:\/} Calculated dust temperature versus distance from
  star. The dust models are labeled as follows: OH5 = Ossenkopf \&
  Henning (1994), LG = Li \& Greenberg (1997), DL = Draine \& Lee
  (1984) and MMP = Mathis et al. (1983).}
  \label{sed.fig}
\end{figure}

\vfill \eject
\begin{figure}[h]
  
\centerline{\hbox{
\psfig{figure=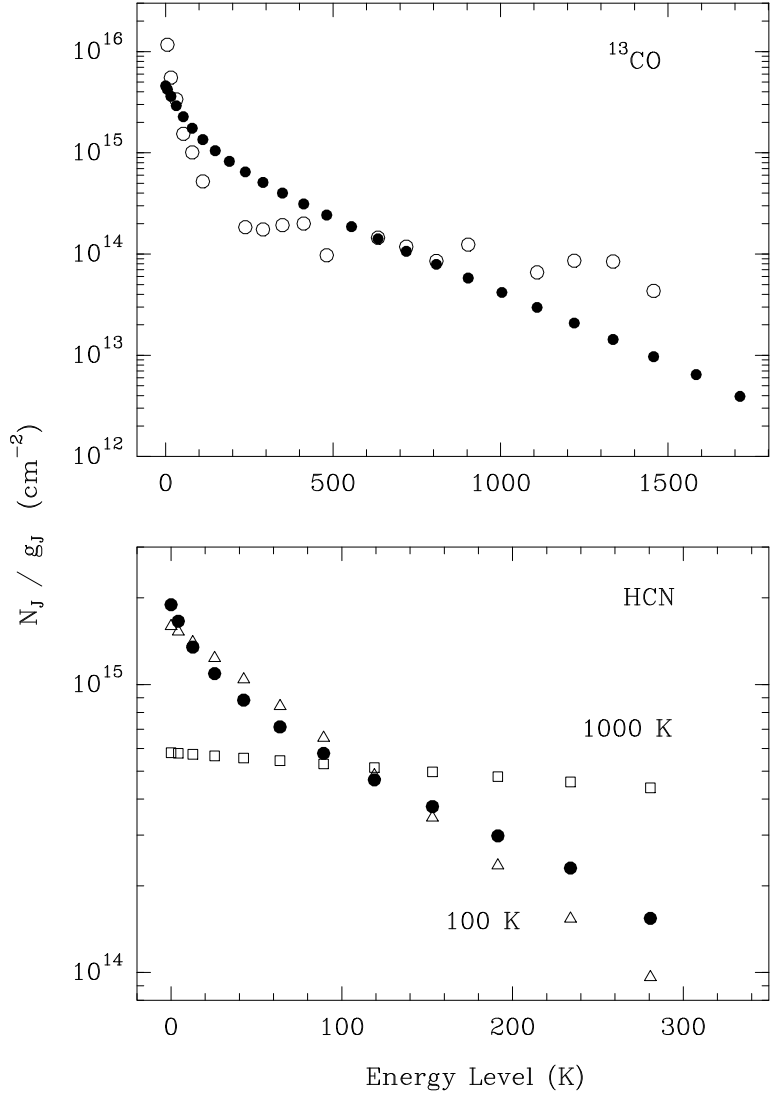,height=18cm,angle=0}
}}

  \caption{Column density per rotational sublevel versus
  level energy for the best fit spherical power law model, denoted by
  filled circles. {\it Top:} \co\ $v=1\gets 0$ band, compared with
  observations by Mitchell et al.  (1989), scaled to their total
  column density of $2.1\times 10^{17}$~\scm. The observational error,
  $2\times 10^{14}/(2J+1)$~\scm, is smaller than the plotting symbol.
  {\it Bottom:} HCN $\nu_2$ band, compared with the populations in the
  case of constant excitation temperatures along the line of sight of
  100~and 1000~K. The value of 1000~K was derived from ISO
  observations by Lahuis \& van Dishoeck (1997). All curves have been
  scaled to N(HCN)$=7.2\times 10^{16}$~\scm. The model does not
  include infrared pumping.}
  \label{rot-synth.fig}
\end{figure}

\vfill \eject
\begin{figure}[h]
  
\centerline{\hbox{
\psfig{figure=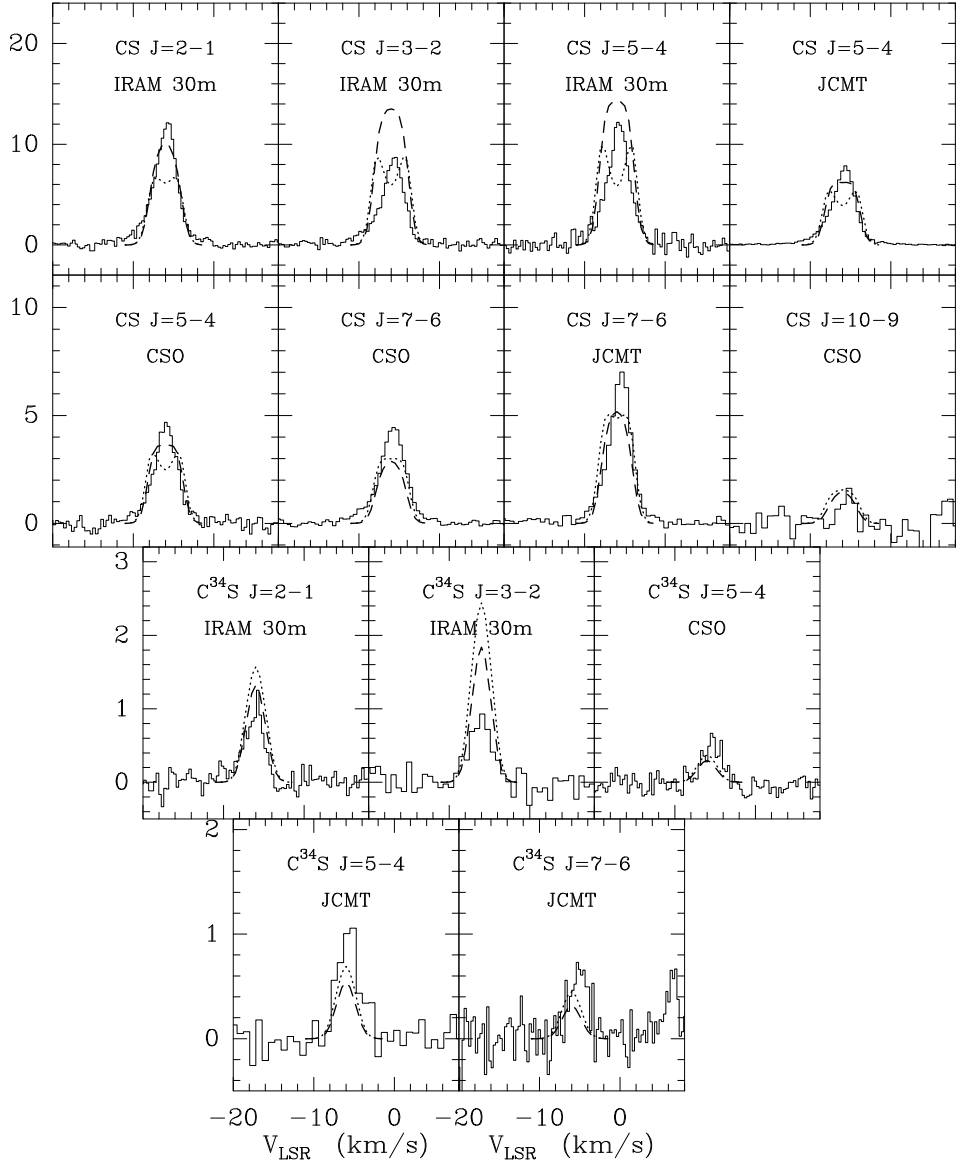,height=18cm,angle=0}
}}

  \caption{Line profiles of CS and \cs, observed with
  the JCMT unless otherwise indicated.  {\it Dotted line:\/} best fit
  spherical power law model with $n($H$_2)=3.5\times 10^4$ (r/30,000
  AU)$^{-1.25}$~\ccm, CS/\htwo\ $=1\times 10^{-8}$.  {\it Dashed
    line:\/} Two-dimensional model with an outflow opening angle of
  $\theta=30^\circ$.}
  \label{mc-cs.fig}
\end{figure}

\vfill \eject
\begin{figure}[h]
  
\centerline{\hbox{
\psfig{figure=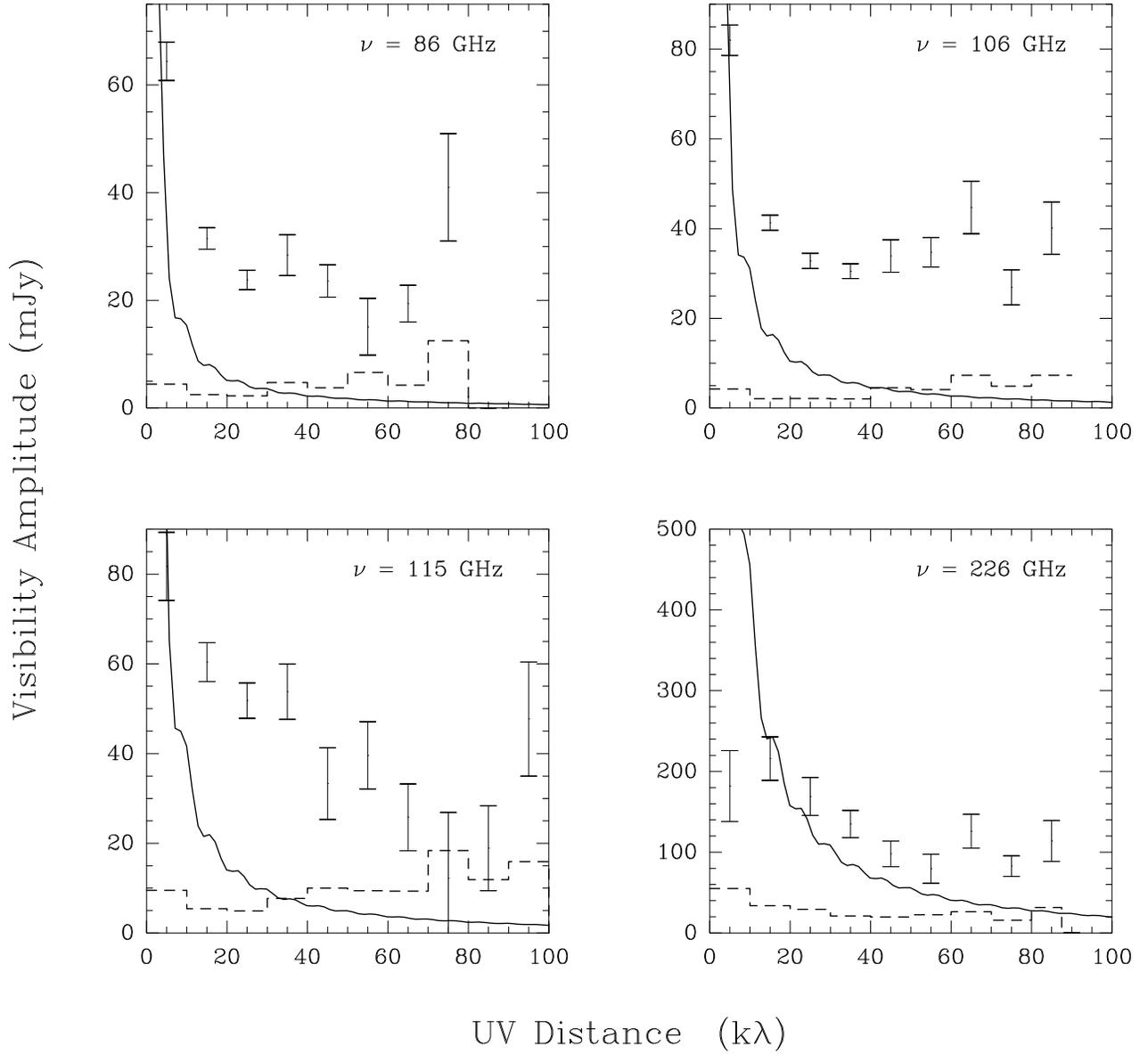,height=18cm,angle=0}
}}

  \caption{Continuum visibilities observed at 86, 106,
  115 and 226 GHz with OVRO, binned in azimuth about the phase center.
  The dashed line indicates the amplitude bias. The solid line is the
  visibility function calculated for the power law model fitted to the
  JCMT continuum data.}
  \label{cvis.fig}
\end{figure}

\vfill \eject
\begin{figure}[h]
  
\centerline{\hbox{
\psfig{figure=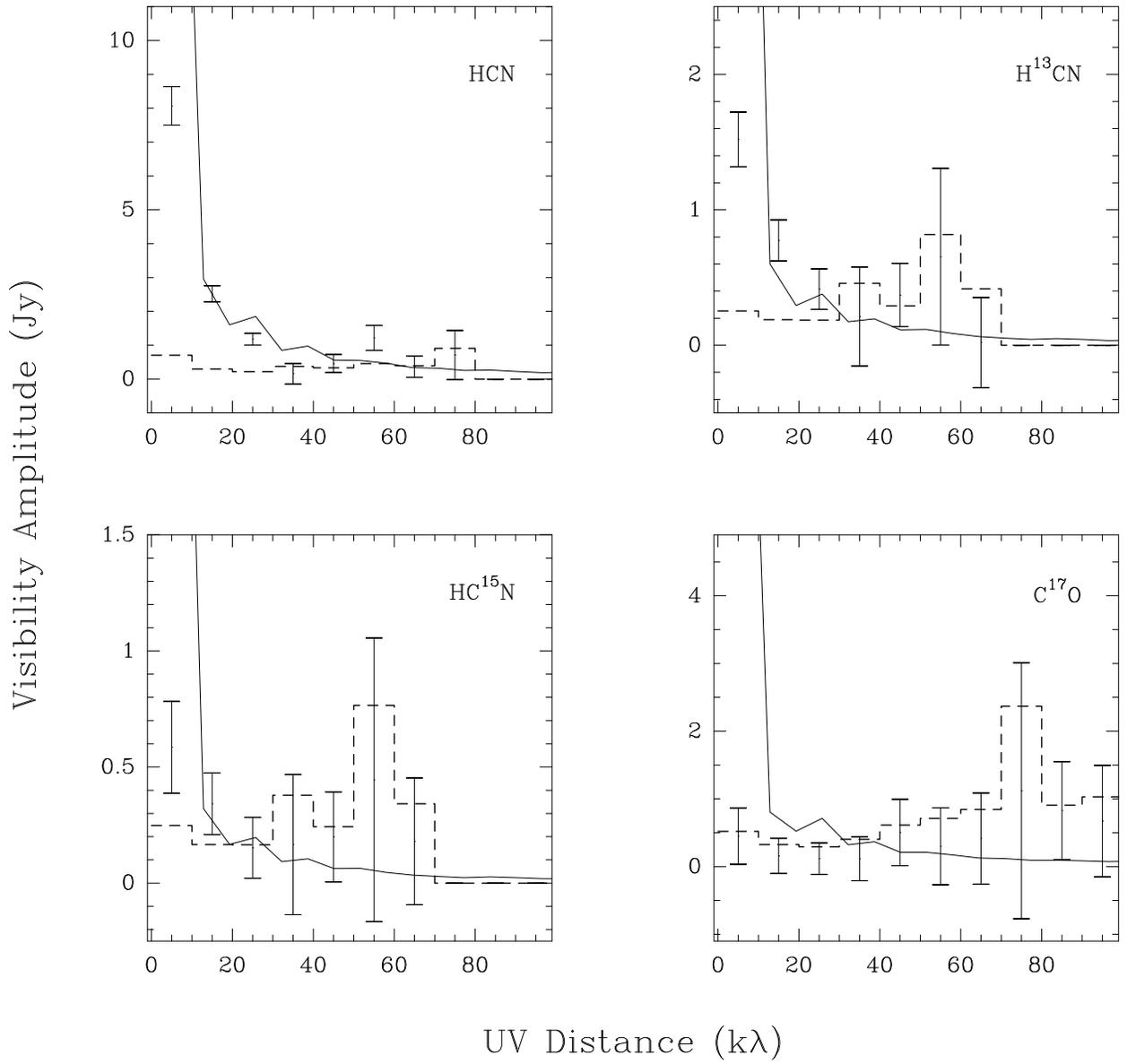,height=18cm,angle=0}
}}

  \caption{As Figure~9, but for the $J=1\to0$ lines
  of HCN, H$^{13}$CN, HC$^{15}$N and C$^{17}$O. The observed emission
  is integrated over 1~\kms.}
  \label{mc-ovro.fig}
\end{figure}

\vfill \eject
\begin{figure}[h]
  
\centerline{\hbox{
\psfig{figure=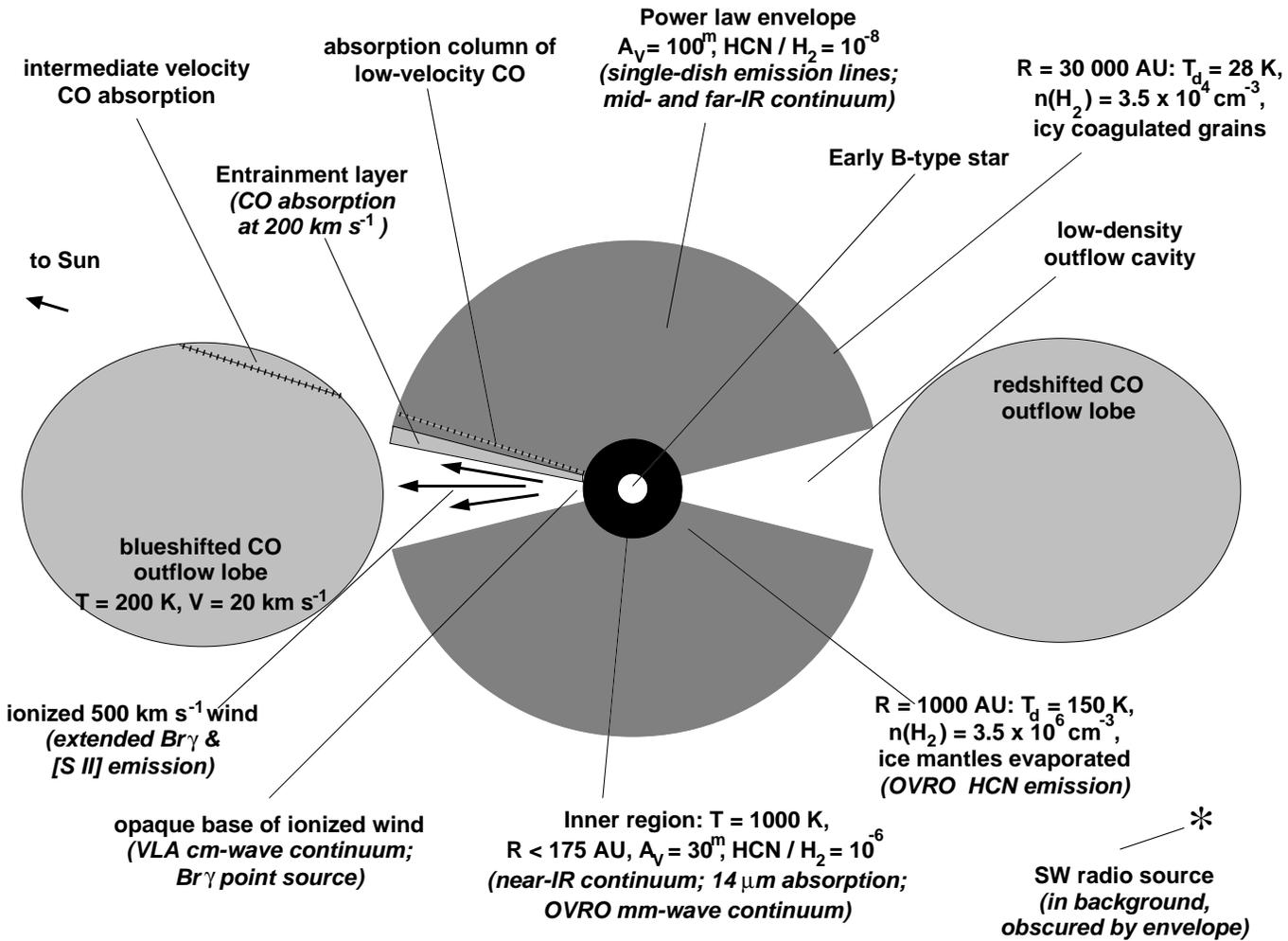,height=18cm,angle=0}
}}

  \caption{Schematic drawing of GL~2591, not to scale,
  with the observational characteristics of the various physical
  components indicated. Velocities are in the rest frame of the
  source; subtract $5.7$~\kms\ to convert to the LSR
  scale.}
  \label{cartoon.fig}
\end{figure}

\end{document}